\documentclass[review]{elsarticle}

\journal{Information Sciences}
\usepackage[pagewise]{lineno}

\usepackage{array,multirow}
\usepackage{caption}
\usepackage{graphicx}
\usepackage{subfig}
\usepackage{amsmath}
\usepackage{algorithm,algorithmic}
\usepackage{geometry}
\newcolumntype{P}[1]{>{\centering\arraybackslash}p{#1}}
\newcolumntype{M}[1]{>{\centering\arraybackslash}m{#1}}
\usepackage{hyperref}
\hypersetup{colorlinks=true,linkcolor=blue,anchorcolor=blue,citecolor=blue,filecolor=blue,urlcolor=blue,bookmarksnumbered=true,pdfview=FitB,linktocpage=true,hypertexnames=false,pdftex}

\usepackage{listings,color}

\begin{document}
	
	\begin{frontmatter}
		
		\title{A novel model for query expansion using pseudo-relevant web knowledge}
		
		\author[]{Hiteshwar Kumar Azad \corref{myca}}
		\ead{hiteshwar.cse15@nitp.ac.in}
		\cortext[myca]{Corresponding Author}
		
		\author[]{Akshay Deepak}
		\ead{akshayd@nitp.ac.in}

		\address{Dept. of Computer Science \& Engineering\\
			National Institute of Technology Patna, India}
%
%
%
%


\begin{abstract}
In the field of information retrieval, query expansion (QE) has long been used as a technique to deal with the fundamental issue of word mismatch between a user's query and the target information. In the context of the relationship between the query and expanded terms, existing weighting techniques often fail to appropriately capture the term-term relationship and term to the whole query relationship, resulting in low retrieval effectiveness. Our proposed QE approach addresses this by proposing three weighting models based on (1) tf-itf, (2) k-nearest neighbor (kNN) based cosine similarity, and (3) correlation score. Further, to extract the initial set of expanded terms, we use pseudo-relevant web knowledge consisting of the top $N$ web pages returned by the three popular search engines namely, Google, Bing, and DuckDuckGo, in response to the original query. Among the three weighting models, \emph{tf-itf} scores each of the individual terms obtained from the web content, \emph{kNN-based cosine similarity} scores the expansion terms to obtain the term-term relationship, and \emph{correlation score} weighs the selected expansion terms with respect to the whole query. The proposed model, called web knowledge based query expansion (WKQE), achieves an improvement of 25.89\% on  the MAP score and 30.83\% on the GMAP score over the unexpanded queries on the FIRE dataset. A  comparative analysis of the WKQE techniques with other related approaches clearly shows significant improvement in the retrieval performance. We have also analyzed the effect of varying the number of pseudo-relevant documents and expansion terms on the retrieval effectiveness of the proposed  model.

\end{abstract}

\begin{keyword}
	\texttt{Query expansion \sep Query reformulation \sep Information retrieval \sep Pseudo relevance feedback \sep Web search \sep Web knowledge}
\end{keyword}

\end{frontmatter}

\section{Introduction}\label{sec1}

Present information retrieval (IR) systems, especially search engines, need to deal with the challenging issue of satisfying a user's needs expressed by short queries. As per recent reports \cite{sta,key}, the most frequent queries consist of one, two, or three words only \cite{azad2019query} -- the same as twenty years ago as reported by Lau and Horvitz \cite{lau1999patterns}. While the users continue to fire short queries, the number of web pages have increased exponentially on the web \cite{merigo2018fifty}. This has increased the ambiguity in  finding the relevant information due to the multiple meanings/senses of the query terms, where indexers and users often do not use the same word. This is also called vocabulary mismatch problem \cite{furnas1987vocabulary}. An effective strategy to resolve this issue is to use query expansion (QE) techniques. Query expansion reformulates the seed query by adding additional relevant terms with similar meaning. The selection of these expansion terms plays a crucial role in QE because only a small subset of the candidate expansion terms is actually relevant to the query \cite{liu2017multi}.  Current commercial search engines do a remarkably good job in interpreting these short queries, however their results can be further improved by using additional external knowledge, obtained by combining their search results, to expand the initial queries.

The data source used for mining the expansion terms plays an important role in QE. A variety of data sources have been explored for mining the expansion terms. These terms may be extracted from an entire document corpus or from a few top-ranked retrieved documents in response to the seed query. A comprehensive survey on data sources used for QE has been provided by Carpineto and Romano \cite{carpineto2012survey} and Azad and Deepak \cite{azad2019query}. Broadly, such sources can be classified into four classes: (i) documents used in retrieval process (e.g., corpus), (ii) hand-built knowledge resources (e.g., WordNet\footnote{https://wordnet.princeton.edu/}, ConceptNet\footnote{http://conceptnet5.media.mit.edu/}, thesaurus, ontology), (iii) external text collections and resources (e.g., Web, Wikipedia, DBpedia), and (iv) hybrid data sources (e.g., combination of two or more data sources).
Among these data sources, external text collections and resources are a popular choice -- even more so in the recent past --  for expanding the user's seed query \cite{lucchese2018efficient,dalton2014entity,bendersky2012effective,yin2009query}. This is because they cover a very diverse range of topics and are curated and regularly updated by a large number of contributors, e.g., Wikipedia\footnote{https://www.wikipedia.org/} and DBpedia\footnote{https://wiki.dbpedia.org/}.  In external text collections and resources, web \cite{lucchese2018efficient,bendersky2012effective}, Wikipedia \cite{dalton2014entity,almasri2013wikipedia}, DBpedia \cite{anand2015empirical}, query logs \cite{yin2009query,cui2002probabilistic}, and anchor texts \cite{dang2010query,kraft2004mining} are the most common and effective data sources for QE. While external text collections like web and Wikipedia cover a diverse range of topics and are regularly updated, a key challenge is to mine these huge data sources for the candidate expansion terms. 

Once a set of candidate expansion terms is determined, the weighting models are used to assess the significance of these terms in relation to the original query. Finally, a few of the candidate expansion terms are selected for inclusion in the expanded query. This selection of the final set of expansion terms is an important factor in QE because only a small set of expansion terms are concretely relevant to the seed query. Here, the selected set of expansion terms should be such that they are well related to the individual terms of the seed query (term-term relationship) as well as to the seed query as a whole. 

This article focuses on query expansion using web knowledge collections from three different search engines namely: Google\footnote{https://www.google.co.in/}, Bing\footnote{https://www.bing.com/}, and DuckDuckGo\footnote{https://duckduckgo.com/}. The Web is the most up-to-date and diversified source of information. This naturally motivates its use as a source for mining expansion terms. Further, popular commercial search engines like Google, Bing, and DuckDuckGo have reasonably mastered the complex technique to mine the Web. They can extract  information that is not only relevant to the original query but also provides a rich set of terms semantically similar to the original query.   Hence, this combination of commercial search engines and the Web appeals as the perfect tool to mine expansion terms. In our proposed model, we have used the pseudo-relevance feedback (PRF) to accumulate the web content of the top $N$ URLs returned by these search engines in response to the seed query. This accumulated web content is then used an external data source to mine candidate expansion terms. 

Pseudo-relevance feedback is an effective strategy in QE to improve the overall retrieval performance of an IR system \cite{lv2011boosting,xu2009query}. It assumes that the top-ranked documents returned in response to the seed query are relevant for mining expansion terms. It uses these relevant `feedback' documents as a source for selecting potentially related terms. In our work, we use the text content of the web pages of the top-ranked URLs returned by search engines in response to the seed query. However, the expansion terms provided by PRF-based methods may not have one-to-one relation with the original query terms, resulting in false expansion terms and causing topic drift.  To address this term-relationship problem, we propose a novel web knowledge based query expansion (WKQE) approach. The proposed WKQE approach uses three modified versions of weighting techniques: (1) tf-itf, (2) k-nearest neighbor (kNN) based cosine similarity, and (3) correlation score. Among these three weighting scores, \emph{tf-itf} and \emph{kNN based cosine similarity} approach scores the expansion terms to determine the term-term relationship, and \emph{correlation score} weighs the selected candidate expansion terms with respect to the whole query.

Experimental results show that the proposed WKQE approach produces consistently better results for a variety of queries (Query ID-126 to 176 in the FIRE dataset)
when compared with the baseline and other related state-of-the-art approaches. We have also analyzed the effect on retrieval effectiveness of the proposed technique by varying the number of pseudo-relevant documents and expansion terms. The experiments were carried on the FIRE dataset using popular weighting models and evaluation metrics.

\subsection{Research contributions}
The research contributions of this paper are follows :

\begin{itemize}
	\item \textbf{Data sources:} This paper presents a novel web knowledge-based query expansion (WKQE) approach that uses the top $N$ pseudo relevant web pages returned by commercial search engines Google, Bing, and DuckDuckGo as data sources. While the Web is the most diversified and up-to-date source of information for mining candidate expansion terms, the commercial search engines are  perfect interfaces to the Web. Based on the literature survey done by us, this seems to be the first use of PRF-based web search results as an external data source for QE.  
	
	\item \textbf{Weighting method:} To appropriately capture the term relationship while weighing the candidate expansion, three modified versions of weighting techniques have been proposed: (1) tf-itf (2) k-nearest neighbor (kNN) based cosine similarity, and (3) correlation score. 
	
	A two-level strategy is employed to select the final set of expansion terms: first, an intermediate set of terms are selected on the basis of term-to-term relationship using \emph{tf-itf} and \emph{kNN based cosine similarity} approach, and then, the final set of terms are selected on the basis of term-to-whole query relationship using \emph{correlation score}.
	
	\item \textbf{Experimental results:}   Experimental evaluation on the Forum for Information Retrieval Evaluation (FIRE) dataset produced an improvement of 25.89\% on  the MAP score and 30.83\% on the GMAP score over the unexpanded queries. Retrieval effectiveness of the proposed technique was also evaluated by varying the number of pseudo-relevant documents and expansion terms. 
	
\end{itemize}

\subsection{Organization}
The rest of this paper is organized as follows. Section \ref{Related Work} discusses related work. Section \ref{Proposed Approach} describes the proposed approach. Section \ref{Experimental Setup} discusses the experimental setup; describing dataset, model parameters,s and evaluation matrices in sub sections.  Section \ref{Evaluation Results} discusses the experimental results. Finally, we conclude in Section \ref{Conclusion}.

%
%
%

\section{Related Work}\label{Related Work}
Query expansion has a long history of literature in the field of information retrieval. It was first coined by Moron et al. \cite{maron1960relevance} in the 1960s for literature indexing and searching in a mechanized library system. In 1971, Rocchio \cite{rocchio1971relevance} brought QE to spotlight through the relevance feedback method and its characterization in a vector space model. While this was the first use of relevance feedback method, Rocchio's method is still used for QE in its original and modified forms. The availability of several standard text collections (e.g., Text Retrieval Conference (TREC)\footnote{http://trec.nist.gov/}, and Forum for Information Retrieval Evaluation (FIRE)\footnote{http://fire.irsi.res.in/}) and IR platforms (e.g., Terrier\footnote{http://terrier.org/} and Apache Lucene\footnote{http://lucene.apache.org/}) have been very instrumental in evaluating the progress in this area in a systematic way. 
Carpineto and Romano \cite{carpineto2012survey} and Azad and Deepak \cite{azad2019query} present state-of-the-art comprehensive surveys on QE. This article focuses on web based QE techniques.

In web based QE techniques, web pages \cite{lucchese2018efficient,bendersky2012effective}, Wikipedia articles \cite{azad2019new,dalton2014entity,almasri2013wikipedia}, query logs \cite{yin2009query,cui2002probabilistic}, and anchor texts \cite{dang2010query,kraft2004mining} are the most common and effective data sources for QE. A common way to utilize the web pages, which are related to the user query, as a data source for query expansion is to use the text snippets of those web pages that are returned by web search engine(s) in response to the user query.  Such a snippet is typically a brief window of text extracted by a search engine around the query term in a web page related to the user query. They are short summaries of the corresponding web pages and are expected to be highly related to the user query, and hence, a rich data source for query expansion. 
Sahmi et al. \cite{sahami2006web} used snippets returned by Google search engine (http://www.google.com/apis) to extract semantically similar terms for QE. For each query, their approach collects snippets from the search engine and represent each snippet as a tf-idf weighted term vector. However, a widely accepted flaw of using snippets is that due to the massive scale of the web and a large number of documents in the result set, only those snippets that are extracted form the top-ranked results can be processed efficiently for improving retrieval performance. For addressing this flaw, Bollegala et al. \cite{bollegala2007measuring} considered the text snippets and page counts returned by the Google search engine. Based on this, they defined various similarity scores  based on page counts for finding relevant expansion terms. Another work based on snippets is by Riezler et al. \cite{riezler2008translating}, where the authors have proposed a query-to-snippet translation model for improving QE. Their proposed model uses user queries and snippets of clicked results to train a machine translation model. This establishes the relationship between query and document space to resolve the lexical gap. Yin et al. \cite{yin2009query} used the search engine query logs, snippets, and search result documents for QE. Their proposed method expresses the search engine query log as a bipartite query-URL graph, where query nodes are connected to the URL nodes by click edges; it reported an improvement of retrieval effectiveness by more than 10\%.

With the fast growing size of the Web and an increasing use of search engines, the abundance of query logs and their ease of use have made them an important source for QE. The query logs usually contains user queries and the corresponding URLs of web pages visited by the user in response to the query results. Here, different users may submit various queries to express the same information-need. Therefore, the query can be expanded by using the wisdom of the crowd. Cui et al. \cite{cui2002probabilistic} used the query logs to extract probabilistic correlations between the query terms and document terms. These correlations are further used for expanding the user's initial query. They extended their work in \cite{cui2003query} to improve upon their  results when compared with QE based on pseudo relevance feedback. One of the advantages of using the query logs is that it implicitly incorporates relevance feedback. On the other hand, it has been shown by White et al. \cite{white2005study} that implicit measurements are relatively good, however, their performance may not be the same for all types of users and search operations.

Based on web search query logs, two types of QE approaches are usually used. The first type extract features from the queries, stored in logs, that are related to the user's original query, with or without making use of their respective retrieval results \cite{huang2003relevant,yin2009query}. In techniques based on the first approach, some use their combined retrieval results \cite{huang2009analyzing}, while some do not (e.g., \cite{huang2003relevant,yin2009query}). In the second type of approach, the features are extracted on relational behavior of queries and retrieval results. For example, Baeza et al. \cite{baeza2011extracting} represent queries in a graph based vector space model (query-click bipartite graph) and analyze the graph constructed using the query logs. Under the second approach, the expansion terms are extracted form several approaches: through user clicks \cite{xue2004optimizing,yin2009query,hua2013clickage}, directly from the clicked results \cite{cui2003query,riezler2007statistical,cao2008context}, the top results from the past query terms entered by the user \cite{fitzpatrick1997automatic,wang2007learn}, and queries related with the same documents \cite{billerbeck2003query,wang2008mining}. However, the second type of approach is more widely used and has been shown to provide better results.

In the context of web-based knowledge, anchor texts can play a role similar to the user's search queries because an anchor text to a page can serve as a brief summary of its content.  Anchor texts were first used by McBryan \cite{mcbryan1994genvl} for associating  hyperlinks with linked pages as well as with the pages in which the anchor texts are found. Kraft and Zien \cite{kraft2004mining} also used anchor texts for QE; their experimental results suggest that anchor texts can be used to improve traditional QE based on query logs. Similarly, Dang and Croft \cite{dang2010query} suggested that anchor text could be an effective alternative to query logs. They demonstrated the effectiveness of QE techniques using log-based stemming through experiments on standard TREC collection dataset.

Another popular approach in the web-based knowledge is the use of Wikipedia articles, titles and hyperlinks (in-link and out-link) \cite{arguello2008document,almasri2013wikipedia,azad2019new} for QE. Wikipedia is the largest encyclopedia available freely on the Web where the articles are regularly updated and new ones are added every day. These features make it an ideal knowledge source for QE. Recently, quite a few research works have used it for QE (e.g., \cite{li2007improving,arguello2008document,xu2009query,aggarwal2012query,almasri2013wikipedia}). Li et al. \cite{li2007improving} performed an investigation using Wikipedia where they retrieved all the articles corresponding to the original query and used them as a source of expansion terms for pseudo-relevance feedback. They observed that for those queries where the general pseudo-relevance feedback failed to improve the query, Wikipedia-based pseudo-relevance feedback improved them significantly. Xu et al. \cite{xu2009query} utilized Wikipedia to categorize the original query into three types: (1) ambiguous queries (queries with terms having more than one potential meaning), (2) entity queries (queries having a specific sense that cover a narrow topic),  and (3) broader queries (queries having neither ambiguous nor specific meaning). They consolidated the expansion terms into the original query and evaluated these techniques using language modeling IR. Al-Shboul and Myaeng \cite{al2014wikipedia} attempted to enrich the initial queries using semantic annotations in Wikipedia articles combined with phrase-disambiguation. Their experiments show better results than the relevance based language model.

\section{Proposed Approach}\label{Proposed Approach}
\begin{figure}[h!]
	\centering 
	\includegraphics[width=11cm, height=14cm]{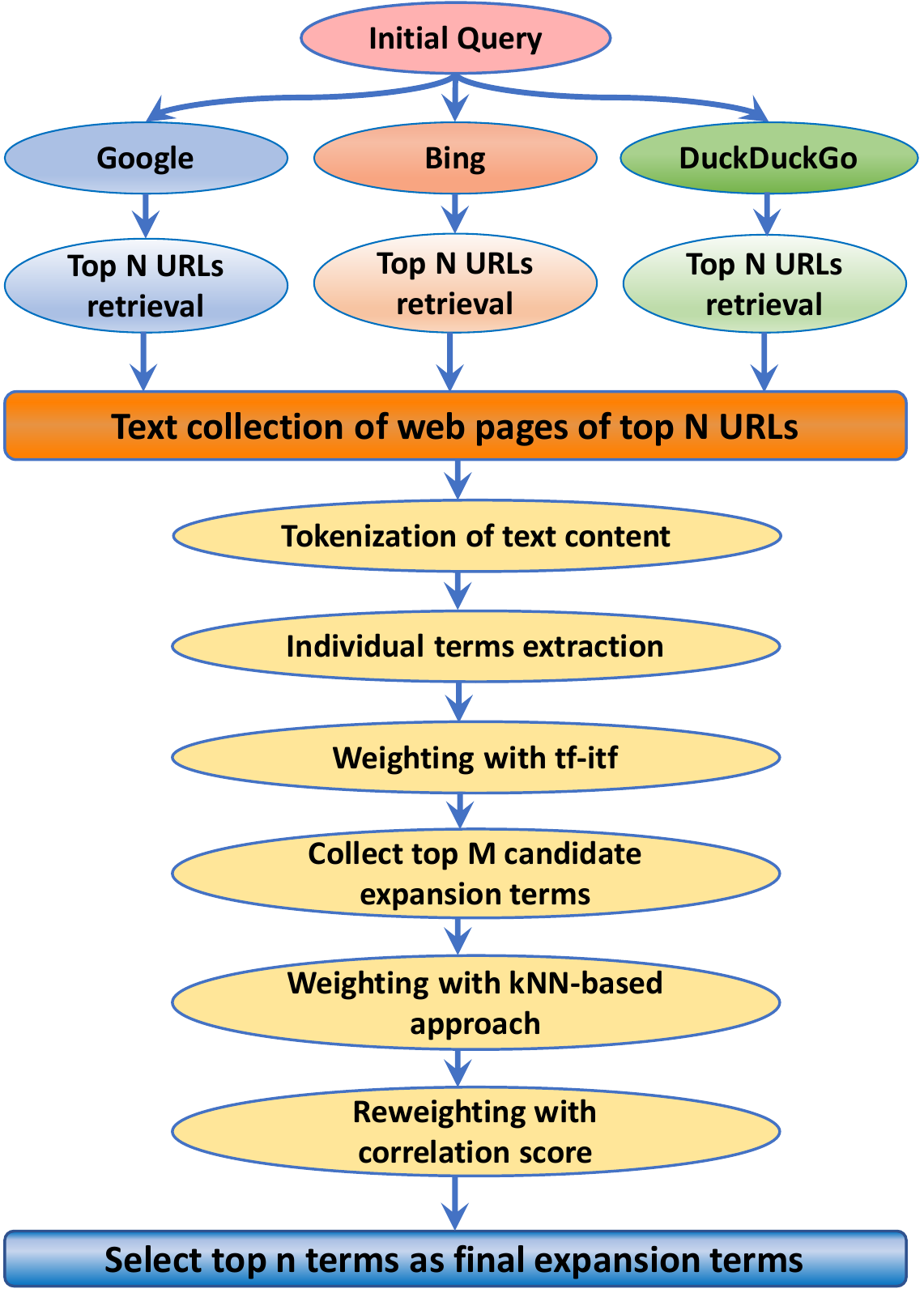}     
	\caption{Steps involved in the proposed approach}
	\label{fig1} 
\end{figure}

During query expansion, the first important decision is the determination of the source for mining candidate expansion terms. 
The top-ranked documents retrieved in response to the initial query appeal be a good source for mining candidate expansion terms. In the context of pseudo relevance feedback, these documents form the set of pseudo-relevant documents. Our proposed approach, called web knowledge based query expansion (WKQE) and shown in Fig. \ref{fig1}, is a pseudo-relevant feedback based technique, where the pseudo-relevant documents consists of the web-pages of the top $N$ URLs returned by three popular search engines namely: Google, Bing, and DuckDuckGo, in response to the initial query. The motivation for doing so has already been discussed earlier in Sec. \ref{sec1}.
The relevant terms found in the collection of these pseudo-relevant documents are used for QE. Sometimes a particular search engine may not provide the result that the user exactly intended. For example, consider the top ten search results on the query term \emph{apple} as returned by the three search engines. While Google and Bing provide results interpreting \emph{apple} only as a company, DuckDuckGo offers results interpreting the query term both as a company as well as the fruit. So, for diversifying the sense of expansion terms we select three popular search engines instead of just one. 

Both term-to-term and term to the whole query relationships are computed for finding the most relevant in the  set of candidate expansion terms. 
To estimate the term-to-term relationship, we weigh the expansion terms with the proposed tf-itf and kNN based cosine similarity score. To estimate the term to the whole query relationship, we weigh the expansion terms with the correlation score.
As shown in Fig. \ref{fig1}, the proposed approach consist of five main steps: (i) retrieval of top $N$ URLs, (ii) text collection and tokenization, (iii) weighting with tf-itf, (iv) weighting with kNN-based approach, and (v) reweighting with correlation score. These steps are described next. 

\subsection{Retrieval of top $N$ URLs }
In order to expand the initial query, first of all we fired the initial query on three popular search engines namely: Google, Bing, and DuckDuckGo. After that, we extracted the web pages corresponding to the top $N$ URLs returned by each of the search engine separately. These web pages act as the set of pseudo-relevant documents for our approach.  When considering the top $N$ URLs, we have excluded the URLs associated with advertising, video, and e-commerce sites. For experimental evaluation, we considered different value of $N$ as 5, 10, 15, 20, 30, 50. However, the proposed model showed the best performance for $N=20$. See Sec. \ref{No of PRD} for more details.

\subsection{Text collection and tokenization}
The entire content of the pseudo-relevant web pages, corresponding to the top $N$ URLs, is not informative. A web page usually has different types of content that are either not related to the topic of the web page or are not useful for the purpose of query expansion. Such items can be:

\begin{itemize}
	\item \emph{Decoration}: Pictures, animations, logos, etc. for attractions or advertising purposes.
	\item \emph{Navigation}: Intra and inter hyperlinks to guide the user in different parts of a web page.  
	\item \emph{Interaction}: Forms to collect user information or provide search services.
	\item Other special words or paragraphs such as copyrights and contact information.
\end{itemize}

In the context of query matching and term weighting, all of the points as mentioned above are considered to be noise and can adversely affect the retrieval performance. For example, if some words from an advertisement embedded in a top-ranked pseudo-relevant web page are chosen as expansion terms, irrelevant documents containing these advertising terms can be ranked highly. Therefore, it is necessary to filter out the noisy information in web pages and retrieve only the contents that are semantically related to the initial query. The Document Object Model (DOM)\footnote{https://www.w3.org/TR/WD-DOM/introduction.html} represents the structure of an HTML page as a tag tree. DOM-based segmentation approaches have been widely used for many years. In our proposed WKQE approach, we use these page segmentation methods where the tags that can represent  useful blocks in a page include $P$ (for paragraph), $H1-H6$ (for heading), $TABLE$ (for table), $UL$ (for list), etc. To extract the relevant text contents from the web pages (HTML and XML documents), we used  Beautiful Soup\footnote{https://www.crummy.com/software/BeautifulSoup/} Python library and web services.

After collecting the text content of the web pages returned by the three search engines, we created a combined corpus $C$ of all the web pages. Then, we tokenized corpus $C$ to identify individual words. For tokenisation  and to remove stop words, we used the Natural Language Toolkit (NLTK)\footnote{https://www.nltk.org/}. The part of speech (POS) tags assigned by NLTK tagger were used to identify phrases and individual words. After the extraction of individual terms, we weighted these individual terms with tf-itf. The weighting of the expansion terms is described next.

\subsection{ Weighting with tf-itf}\label{tf}
We weighted the individual terms with tf-itf, which is a modified version of tf-idf. This weight is a statistical measure used to evaluate how important a word in a corpus is. The term frequency (tf) measures how frequently a term occurs in a corpus. While computing tf, all terms are considered equally important. However, it is known that certain terms in a corpus, such as ``is", ``of", and ``that", may appear a lot of times but have little importance. Thus we need to weight down such frequent terms while scale up the importance of the rare ones. This is achieved  by computing the inverse term frequency (itf). We weight the individual terms with tf-itf scoring function as follows:
\begin{equation}\label{eqn1}
\begin{split}
Score(t_i) &= tf(t_i, C) \:.\: itf(t_i ,C) \\
& =tf(t_i, C) \:.\: log \frac{T}{|t_i|}
\end{split}
\end{equation}
where:\\ $ tf(t_i, C)$ denotes the term frequency of term $t_i$ in the entire corpus $C$, \\ $itf(t_i, C)$ denotes  the inverse term frequency of $t_i$  in the entire corpus $C$,\\ $T$ is the number of terms in the entire corpus $C$, and\\ $|t_i|$ is the number of times term $t_i$ appears in the entire corpus $C$.

After assigning this score to each of the terms in the corpus, we ranked these terms according to the scoring value. Then, we selected the top $M$ individual terms as intermediate candidate expansion terms. After this, these intermediate candidate expansion terms were re-weighted with the kNN-based cosine similarity score, which is described next.

\subsection{ Weighting with kNN-based approach}\label{KNN}

The kNN-based approach weights the intermediate candidate expansion terms with cosine similarity and selects the top $k$ nearest neighbor candidate expansion terms.
It establishes a term-term correlation among the candidate expansion terms so that the most relevant expansion terms can be chosen. 
The proposed kNN-based approach is an extension of the technique given by Roy et al. \cite{roy2016using}. Here, instead of computing the nearest neighbors for each query term, we computed the nearest neighbors of each intermediate candidate expansion term extracted in response to the original query. The highly similar neighbors have comparatively lower drift in terms of similarity  than the terms occurring later in the list. Since the most similar terms are the strongest contenders for becoming the expansion terms, it can be assumed that these terms are also similar to each other, in addition to being similar to the query term. Based on this, we use an iterative process for  sorting the expansion terms described in Algorithm \ref{alg:KNN}.

\begin{algorithm}[t]
	\caption{k-Nearest Neighbors}\label{alg:KNN}
	\begin{algorithmic}[1]
		\renewcommand{\algorithmicrequire}{\textbf{Input:}}
		\renewcommand{\algorithmicensure}{\textbf{Output:}}
		\REQUIRE $C_{exp}\gets$ Set of intermediate candidate expansion terms sorted using Eq.\ref{eqn1}.
				\\ \hspace{6mm}
				 $k\gets$ Number of nearest neighbor candidate expansion terms to be returned. 
				\\  \hspace{6mm}
				$l\gets$ Number of terms terms to be dropped during each iteration. 
		\ENSURE  $NN \gets$ Set of iteratively added nearest neighbor candidate expansion terms.
		\STATE \textit{Initialisation}: $NN \gets \{\emptyset\}$; $r \gets 5$ \hspace{4cm} \# \textit{$r \gets$ Number of iterations.}\label{init}
				\STATE Select $t \in C_{exp}$ having maximum score  \label{line1.1}
				\WHILE  {$r > 0$}
				\STATE $NN \gets NN \cup \{t\}$ \hspace{8cm} \# \textit{Add $t$ to $NN$.}\label{line4}
				\STATE $C_{exp} \gets C_{exp}-\{t\}$ \hspace{5.4cm} \# \textit{Remove the term $t$ from $C_{exp}$.}\label{line5}
				\STATE Sort $C_{exp}$ w.r.t. $t$ using cosine similarity score of Eqn. (\ref{eqn3})\label{line3}
				\STATE Select $t \in C_{exp}$ having maximum score
				\STATE Select $C_{l}\subset C_{exp}$ as the set of $l$ least scoring terms in $C_{exp}$ 
		
				\STATE $C_{exp} \gets C_{exp}- C_{l}$ \hspace{2cm} \# \textit{Remove the set of $l$ least neighbors terms from $C_{exp}$.}\label{line6}
				\STATE Select $t \in C_{exp}$ having maximum score  
				\STATE $r \gets r-1$
				\ENDWHILE 
				\STATE Select $C_{k-r}\subset C_{exp}$ as the set of $k$ highest scoring terms in $C_{exp}$ \label{line7}
				\STATE $NN \gets NN \cup C_{k-r}$ \label{line8}
				\STATE \textbf{return} $NN$ \hspace{3cm} \# \textit{Final set of nearest neighbor candidate expansion terms.} 
			\end{algorithmic}
		\end{algorithm}

Algorithm \ref{alg:KNN} takes as input the sorted (using Eq. \ref{eqn1}) set of candidate expansion terms obtained from the previous step, denoted as $C_{exp}$. First, the expansion term  having maximum similarity score in $C_{exp}$ is identified as $t$ (line \ref{line1.1} in Algorithm \ref{alg:KNN}). Term $t$ acts as the nearest neighbor for the first iteration of the iterative process (lines \ref{line3}-\ref{line6}). At the start of each iteration, the nearest neighbor $t$ is added to the set of nearest neighbors $NN$, which is initialized as empty (line \ref{init}), and removed from the set of intermediate candidate expansion terms $C_{exp}$ (lines \ref{line4} and \ref{line5}). The terms in $C_{exp}$ are then sorted based on their proximity with $t$ computed on the basis of cosine similarity. The cosine similarity between two terms $t_k$ and $t_i$ is given as:	
\begin{equation}\label{eqn3}
\begin{split}
Sim_{cosine}(t_k, t_i) &= \frac{c_{t_k, t_i}}{\sqrt{\sum_{d_j}^{}w^2 _{t_k, j}\:.\: \sum_{d_j}^{}w^2 _{t_i, j}}} \\
& =\frac{\sum_{d_j}^{}w _{t_k, j}\:.\: w _{t_i, j}}{\sqrt{\sum_{d_j}^{}w^2 _{t_k, j}\:.\: \sum_{d_j}^{}w^2 _{t_i, j}}}
\end{split}
\end{equation} 
where:\\ $ c_{t_k, t_i}$ denotes the correlation score of term $t_k$ with $t_i$ in the entire corpus $C$ and \\ $w _{t_i, j}\:(w _{t_k, j})$ is the weight of term $t_i$ ($t_k)$ in the document $d_j$ (as returned by one of the three search engines).  $w_{t_i, j}$ is computed as ($w_{t_k, j}$ is similarly defined):
\begin{equation}
\begin{split}
w_{t_i, j} &= tf(t_i, j) \:.\: itf(t_i ,j) \\
& =tf(t_i, j) \:.\: log \frac{T}{|DT_j|}
\end{split}
\end{equation}
where:\\ $ tf(t_i, j)$ denotes the term frequency of $t_i$ in the document $d_j$. \\ $itf(t_i, j)$ is the inverse term frequency of $t_i$  in the  document $d_j$.\\ $|DT_j|$ is the number of distinct terms in the document $d_j$, and\\ $T$ is the number of terms in the entire collection.

Then, the $l$ least similar neighbors are removed from the intermediate set of candidate expansion terms $C_{exp}$. This completes the first iteration. This iterative process is executed for $r$ iterations. In each iteration, the nearest neighbors list is rearranged on the basis of the nearest neighbor obtained from the previous iteration and a set of $l$ least similar neighbors is eliminated. Essentially, by following the above procedure, we are compelling the nearest neighbors to be similar to each other in addition to being similar to the original query. 

A high value of $r\geq 10$ may lead to query drift, while a low value  $r\leq 2$ essentially performs similar to the initial set of intermediate candidate expansion terms. In our proposed method we chose $r=5$ as the number of iterations. 
Finally, at the end of $r$ iterations, the top $k$ nearest neighbor candidate expansion terms are returned as the final set of nearest neighbor candidate expansion terms (lines \ref{line7} and \ref{line8}). Now these intermediate candidate expansion terms are reweighted using correlation score (described next) and the top $n$ terms are chosen as the final set of expansion terms.

\subsection{Reweighting with correlation score}\label{reweighting}
So far a set of candidate expansion terms have been obtained, where each expansion term is strongly connected to the other individual candidate expansion terms. These terms have been allocated weights using tf-itf and the kNN-based approach. Things done so far resolve the issue of semantic similarity between term-to-term relationship. However, this may not accurately reflect the relationship of an expansion term to the query as a whole. For example, while the word ``program" may be highly associated with the word ``computer", use of this association for selecting candidate expansion terms may work well for some queries such as ``Java program" and ``application program" but not for others such as ``space program", ``TV program", and ``government program". This problem has been analyzed in Bai et al. \cite{bai2007using}. To address this language ambiguity problem, we use a weighting scheme called correlation score. A similar approach has been suggested in Xu and Croft \cite{xu1996query}, Cui et al. \cite{cui2003query}, Sun et al. \cite{sun2006mining}, and, Azad and Deepak \cite{azad2019new}. This approach extends the term-to-term association methods described previously  in Sections \ref{tf} and \ref{KNN}. In this approach we used term-to-term correlation and computed the correlation score of a given candidate expansion term $t_i$ to each query term. We  then combined the found score to find the correlation to the initial query $Q$.

The correlation score is defined as follows. Let $Q$ be the original query having  individual terms $q_k$s and let $t_i$ be a candidate expansion term. Then, the correlation score of $t_i$ with $Q$, denoted $C_{t_i, Q}$, is computed as:

\begin{equation}
\begin{split}
C_{t_i, Q} &= \frac{1}{|Q|} \:.\: \sum_{q_k \in Q }^{}c _{t_i, q_k} \\
& =\frac{1}{|Q|} \:.\:\sum_{q_k \in Q }\sum_{d_j }^{}w _{t_i, j}\:.\: w _{q_k, j}
\end{split}
\end{equation}

where:\\ $ c_{t_i, q_k}$ is the correlation (similarity) score between the candidate expansion term $t_i$ and the query term $q_k$ and \\ $w _{t_i, j}\:(w _{q_k, j})$ is the weight of term $t_i$ ($q_k$) in the document $d_j$.\\
The weight of the candidate expansion term $t_i$ in the document $d_j$, denoted $w_{t_i, j}$ ($w_{q_k, j}$ is similarly defined), is computed as: 
\begin{equation}
\begin{split}
w_{t_i, j} &= tf(t_i, j) \:.\: itf(t_i ,j) \\
& =tf(t_i, j) \:.\: log \frac{T}{|DT_j|}
\end{split}
\end{equation}
where:\\ $ tf(t_i, j)$ denotes the term frequency of $t_i$ in the document $d_j$. \\ $itf(t_i, j)$ is the inverse term frequency of $t_i$  in the  document $d_j$.\\ $|DT_j|$ is the number of distinct terms in the document $d_j$, and\\ $T$ is the number of terms in the entire collection.

After assigning the correlation score to candidate expansion terms, we collect the top $n$ terms as the final set of candidate expansion terms.

\section{Experimental Setup}\label{Experimental Setup}
This section discusses the evaluation of the proposed WKQE approach. Section \ref{dataset} describes the dataset used, followed by discussion of model parameters in Sec. \ref{parameters}. Section \ref{Evaluation Metrics} describes the evaluation metrics used.

\subsection{Dataset}\label{dataset}
To evaluate the proposed technique, the experiments were carried out on a large number of queries (or, Topics) from the well-known benchmark  FIRE\footnote{http://fire.irsi.res.in/} dataset. As queries in the real life are short, we have used only the title field of all queries. Table \ref{tab1} shows the details of the FIRE test collections used in our investigation. These datasets consist of a very large set of documents on which IR is done, a set of queries (called topics), and the right answers (called relevance judgments) stating relevance of  documents to the corresponding topic(s). Specifically, the FIRE ad hoc dataset consists of a large collection of newswire articles from two sources namely The Telegraph\footnote{https://www.telegraphindia.com/} and BDnews24\footnote{https://bdnews24.com/} provided by Indian Statistical Institute Kolkata, India. 
\begin{table}[!h]
	\centering
	\caption{Details of experimental corpora \label{tab1}}{
		
		\begin{tabular}{ | M{2.2cm} | M{2.2cm} | M{2cm} | M{2cm} | M{2cm} | }
			\hline

			\textbf{Corpus} & \textbf{ Disk / Source} & \textbf{Size} & \textbf{\# of docs} & \textbf{Query ID} \\ \hline 
			FIRE  & FIRE 2011 (English) & 1.76 GB & 3,92,577 & 126 - 175 \\ \hline
			
	\end{tabular}}
\end{table}	 

\subsection{Model parameters}\label{parameters}
In order to investigate the optimal value of parameters, we have explored different numbers of top ranked feedback documents, i.e., $N\in \{5, 10, 15, 20, 30, 50\} $, from the three search engines (Google, Bing, and DuckDuckGo). We found that our proposed model performed best for the top 20 feedback documents, hence, we chose the top 20 feedback documents to expand the initial query in our experiments. We also explored different numbers of expansion terms, i.e., $M \in \{ 5, 10, 15, 20, 25, 30, 50\}$, to evaluate the model performance. Our proposed model performed best for the top 15 candidate expansion terms, hence, we chose the top 15 candidate expansion terms to reformulate the original query in our experiments. 

We used the TERRIER\footnote{http://terrier.org/} retrieval system for our all experimental evaluation. We used the title field of the topics in the test collections. For indexing the documents, first stopwords were removed and then Porter's Stemmer was used for stemming the documents. All experimental evaluations are based on the unigram word assumption, i.e., all documents and queries in the corpus are indexed using single terms. We did not use any phrase or positional information. To compare the effectiveness of our expansion technique, we used the following weighting models:  the BM25 model of Okapi \cite{robertson1996okapi}, IFB2 a probabilistic divergence from randomness (DFR) model \cite{amati2002probabilistic}, Laplace's law of succession I(n)L2 \cite{good1965estimation}, Divergence from Independence model DPH \cite{amati2008fub}, Log-logistic DFR model LGD \cite{clinchant2010information}, and Standard tf-idf model. The Parameters for these models were set to the default values in TERRIER.

\subsection{Evaluation Metrics}
\label{Evaluation Metrics}

We evaluated the results on standard evaluation metrics: MAP (Mean Average Precision), GM\_MAP (Geometric Mean Average Precision), F-Measure, P@5, P@10, P@20, P@30 (P@$x$ denotes precision at top $x$ ranks), bpref (binary preference), and the overall recall (number of relevant documents retrieved). The evaluation metric MAP reflects the overall performance of the system, P@5 measures the precision over the top 5 documents retrieved, and bpref measures a preference relation about how many documents judged as relevant are ranked before the documents judged as irrelevant. Additionally, we have reported the percentage improvement in MAP over the baseline (non expanded query) for each expansion method and other related methods. We have also investigated the retrieval effectiveness of the proposed technique with number of expansion terms.

\section{Evaluation Results}\label{Evaluation Results}
The objective of our experiments is to explore the effectiveness of the proposed Web Knowledge based QE (WKQE) approach and to compare it with the baseline as well as  existing state-of-the-art methods on popular weighting models and evaluation metrics. The proposed WKQE approach can be categorised into seven different techniques, namely GQE (Google-based query expansion), BQE (Bing-based query expansion), DQE (DuckDuckGo-based query expansion), GBQE (Google-Bing-based query expansion), GDQE (Google-DuckDuckGo-based query expansion), BDQE (Bing-DuckDuckGo-based query expansion), and GBDQE (Google-Bing-DuckDuckGo-based query expansion). We compared each approach with the baseline (unexpanded query) as well as with the existing state-of-the-art methods. We found that the proposed GBDQE approach gives the best results compared to other approaches. 

Table \ref{QE using Google alone}--\ref{QE using GoogleBingDuckDuckGo} shows the comparative retrieval performance of the proposed WKQE approach using popular weighting models and with respect to evaluation metrics, namely MAP, GM\_MAP, P@10, P@20, P@30, and relevant return. The table shows that the proposed WKQE approach  is compatible with the existing popular weighting models and significantly improves upon the retrieval performance over the unexpanded query. It also shows the relative percentage improvements (within parentheses) of various standard evaluation metrics measured against the unexpanded query. In all the cases, the MAP improvement is more than 4.84\% and the maximal improvement achieved by our proposed QE technique is up to 25.89\%. Based on the results presented in Tables \ref{QE using Google alone}--\ref{QE using GoogleBingDuckDuckGo}, we can say that in the context of all evaluation parameters, the proposed QE technique performs well with all weighting models with respect to the baseline approach of unexpanded query.

Table \ref{QE using Google alone} presents the comparative analysis of QE using Google alone (GQE) on different popular weighting models with top 15 expansion terms. In the best case, GQE improved the MAP up to 22.35\% and GM\_MAP up to 29.08\%. The improvement over the precision reaches its its best value of 25.14\% on the top 10 feedback documents. Overall, GQE produced the best results among the individual query expansion techniques (i.e., GQE, BQE, and DQE), even better than some of the combined QE techniques (e.g., BDQE and GDQE).

\begin{table}[!h]
	\centering
	\caption{Comparison of QE using Google alone (GQE) on popular models with top 15 expansion terms on the FIRE Dataset. The best result for each method have been highlighted in bold. \label{QE using Google alone}}{
		\begin{tabular}{ |p{1.4cm}||p{1.5cm}|p{1.5cm}|p{1.5cm}|p{1.5cm}|p{1.5cm}|p{1.5cm}|  }
			\hline
			\multicolumn{7}{|c|}{\textbf{Model Performance Without Query Expansion}} \\
			\hline
			Method & MAP & GM\_MAP & P@10 & P@20 & P@30 & \#rel\_ret \\
			\hline
			IFB2  & 0.2765 & 0.1907 & 0.3660 & 0.3560 & 0.3420 & 2330\\
			LGD & 0.2909 & 0.1974 & 0.4100 & 0.3710 & 0.3420 & 2309\\
			I(n)L2& 0.2979 & 0.2023 & 0.4280 & 0.3900 & 0.3553 &  2322\\
			DPH & 0.3133 & 0.2219 & 0.4540 & 0.4040 & 0.3653 & 2338\\
			TF\_IDF & 0.3183 & 0.2261 & 0.4560 & 0.4010 & 0.3707 & 2340\\
			BM25 & 0.3163 & 0.2234 & 0.4600 & 0.3970 & 0.3660 &  2343\\
			\hline \hline
			\multicolumn{7}{| c |}{\textbf{Model Performance With QE using Google alone}}\\ \hline
			IFB2  & 0.3383 \textbf{($\uparrow$22.35\%)}& 0.2427 ($\uparrow$27.27\%) & 0.4580 \textbf{($\uparrow$25.14\%)} & 0.4240 \textbf{($\uparrow$19.10\%)} & 0.4013 ($\uparrow$17.34\%) & 2558 ($\uparrow$9.79\%)\\
			
			LGD & 0.3530 ($\uparrow$21.34\%) & 0.2548 \textbf{($\uparrow$29.08\%)} & 0.4680 ($\uparrow$14.15\%) & 0.4330 ($\uparrow$16.71\%) & 0.4080 \textbf{($\uparrow$19.30\%)}& 2550 ($\uparrow$10.44\%)\\
			I(n)L2& 0.3492 ($\uparrow$17.22\%) & 0.2518 ($\uparrow$24.47\%) &  0.4780 ($\uparrow$11.68\%) & 0.4270 ($\uparrow$9.49\%) & 0.4000 ($\uparrow$12.58\%) &  2566 \textbf{($\uparrow$10.51\%)}\\
			DPH & 0.3622 ($\uparrow$15.6\%) & 0.2650 ($\uparrow$19.42\%) & 0.4820 ($\uparrow$6.17\%) & 0.4420 ($\uparrow$9.4\%) & 0.4200 ($\uparrow$14.97\%)& 2572 ($\uparrow$10.0\%)\\
			TF\_IDF & 0.3573 ($\uparrow$12.25\%) & 0.2620 ($\uparrow$15.88\%) & 0.4580 ($\uparrow$0.44\%) & 0.4360 ($\uparrow$8.73\%) & 0.4133 ($\uparrow$11.49\%)& 2556 ($\uparrow$9.23\%)\\
			BM25 & 0.3549 ($\uparrow$12.2\%) & 0.2601 ($\uparrow$16.43\%) &  0.4620 ($\uparrow$0.43\%) & 0.4350 ($\uparrow$9.57\%) & 0.4060 ($\uparrow$10.93\%) &  2555 ($\uparrow$9.05\%)\\
			\hline
	\end{tabular}}
\end{table}

Table \ref{QE using Bing alone} shows the corresponding  comparative analysis of QE using Bing alone (BQE).
Here, MAP is improved up to 21.16\% and GM\_MAP up to 27.63\%. For BM25 weighting model in particular, it has been shown that the value of precision for top 10 retrieval (P@10) is reduced by 1.32\%. The main reason behind the P@10 reduction is the low availability of the expansion terms in the top 10 retrieval documents. Overall BQE showed  better retrieval performance in comparison to DQE, BDQE, and GDQE.

\begin{table}[!h]
	\centering
	\caption{Comparison of QE using Bing alone (BQE) on popular models with top 15 expansion terms on the FIRE Dataset. The best result for each method have been highlighted in bold.\label{QE using Bing alone}}{
		\begin{tabular}{ |p{1.4cm}||p{1.5cm}|p{1.5cm}|p{1.5cm}|p{1.5cm}|p{1.5cm}|p{1.5cm}|  }
			\hline
			\multicolumn{7}{|c|}{\textbf{Model Performance Without Query Expansion}} \\
			\hline
			Method & MAP & GM\_MAP & P@10 & P@20 & P@30 & \#rel\_ret \\
			\hline
			IFB2  & 0.2765 & 0.1907 & 0.3660 & 0.3560 & 0.3420 & 2330\\
			LGD & 0.2909 & 0.1974 & 0.4100 & 0.3710 & 0.3420 & 2309\\
			I(n)L2& 0.2979 & 0.2023 & 0.4280 & 0.3900 & 0.3553 &  2322\\
			DPH & 0.3133 & 0.2219 & 0.4540 & 0.4040 & 0.3653 & 2338\\
			TF\_IDF & 0.3183 & 0.2261 & 0.4560 & 0.4010 & 0.3707 & 2340\\
			BM25 & 0.3163 & 0.2234 & 0.4600 & 0.3970 & 0.3660 &  2343\\
			\hline \hline
			\multicolumn{7}{| c |}{\textbf{Model Performance With QE using Bing alone}}\\ \hline
			IFB2  & 0.3350 \textbf{($\uparrow$21.16\%)}& 0.2434 \textbf{($\uparrow$27.63\%)} & 0.4360 \textbf{($\uparrow$19.12\%)} & 0.4140 ($\uparrow$16.29\%) & 0.4013 ($\uparrow$17.34\%) & 2466 \textbf{($\uparrow$5.84\%)}\\
			
			LGD & 0.3428 ($\uparrow$17.84\%) & 0.2376 ($\uparrow$20.36\%) & 0.4520 ($\uparrow$10.24\%) & 0.4360 \textbf{($\uparrow$17.52\%)} & 0.4027 \textbf{($\uparrow$17.75\%)} & 2436 ($\uparrow$5.50\%)\\
			I(n)L2& 0.3355 ($\uparrow$12.62\%) & 0.2342 ($\uparrow$15.77\%) &  0.4480 ($\uparrow$4.67\%) & 0.4170 ($\uparrow$6.92\%) & 0.3993 ($\uparrow$12.38\%) &  2431 ($\uparrow$4.69\%)\\
			DPH & 0.3497 ($\uparrow$11.62\%) & 0.2477 ($\uparrow$11.63\%) & 0.4640 ($\uparrow$2.20\%) & 0.4440 ($\uparrow$9.90\%) & 0.4133 ($\uparrow$13.14\%)& 2451 ($\uparrow$4.83\%)\\
			TF\_IDF & 0.3482 ($\uparrow$9.39\%) & 0.2472 ($\uparrow$9.33\%) & 0.4660 ($\uparrow$2.19\%) & 0.4410 ($\uparrow$9.97\%) & 0.4173 ($\uparrow$12.57\%)& 2450 ($\uparrow$4.70\%)\\
			BM25 & 0.3434 ($\uparrow$8.57\%) & 0.2421 ($\uparrow$8.37\%) &  0.4540 ($\downarrow$1.32\%) & 0.4340 ($\uparrow$9.32\%) & 0.4073 ($\uparrow$11.28\%) &  2442 ($\uparrow$4.22\%)\\
			\hline
	\end{tabular}}
\end{table}

Table \ref{QE using DuckDuckGo alone} shows the corresponding results for DuckDuckGo-based QE (DQE). The DQE improved the MAP up to 13.03\% and GM\_MAP up to 13.06\%. The DQE showed best retrieval effectiveness with the top 10 expansion terms instead of the top 15 expansion terms used in the other proposed expansion techniques. Overall it showed less improvement when compared to the other proposed expansion techniques. 

\begin{table}[!h]
	\centering
	\caption{Comparison of QE using DuckDuckGo alone (DQE) on popular models with top 10 expansion terms on the FIRE Dataset. The best result for each method have been highlighted in bold. \label{QE using DuckDuckGo alone}}{
		\begin{tabular}{ |p{1.4cm}||p{1.5cm}|p{1.5cm}|p{1.5cm}|p{1.5cm}|p{1.5cm}|p{1.5cm}|  }
			\hline
			\multicolumn{7}{|c|}{\textbf{Model Performance Without Query Expansion}} \\
			\hline
			Method & MAP & GM\_MAP & P@10 & P@20 & P@30 & \#rel\_ret \\
			\hline
			IFB2  & 0.2765 & 0.1907 & 0.3660 & 0.3560 & 0.3420 & 2330\\
			LGD & 0.2909 & 0.1974 & 0.4100 & 0.3710 & 0.3420 & 2309\\
			I(n)L2& 0.2979 & 0.2023 & 0.4280 & 0.3900 & 0.3553 &  2322\\
			DPH & 0.3133 & 0.2219 & 0.4540 & 0.4040 & 0.3653 & 2338\\
			TF\_IDF & 0.3183 & 0.2261 & 0.4560 & 0.4010 & 0.3707 & 2340\\
			BM25 & 0.3163 & 0.2234 & 0.4600 & 0.3970 & 0.3660 &  2343\\
			\hline \hline
			\multicolumn{7}{| c |}{\textbf{Model Performance With QE using DuckDuckGo alone}}\\ \hline
			IFB2  & 0.3066 ($\uparrow$10.89\%) & 0.2156 \textbf{($\uparrow$13.06\%)} & 0.4340 \textbf{($\uparrow$18.58\%)} & 0.3980 \textbf{($\uparrow$11.80\%)} & 0.3747 ($\uparrow$9.56\%) & 2491 ($\uparrow$6.90\%)\\
			
			LGD & 0.3288 \textbf{($\uparrow$13.03\%)} & 0.2229 ($\uparrow$12.92\%) & 0.4560 ($\uparrow$11.22\%) & 0.4110 ($\uparrow$10.78\%) & 0.3907 \textbf{($\uparrow$14.24\%)} & 2470 ($\uparrow$6.97\%)\\
			I(n)L2& 0.3297 ($\uparrow$10.67\%) & 0.2293 ($\uparrow$13.35\%) &  0.4480 ($\uparrow$4.67\%) & 0.4240 ($\uparrow$8.72\%) & 0.3873 ($\uparrow$9.00\%) &  \textbf{2495 ($\uparrow$7.45\%)}\\
			DPH & 0.3380 ($\uparrow$7.88\%) & 0.2359 ($\uparrow$6.31\%) & 0.4580 ($\uparrow$0.88\%) & 0.4290 ($\uparrow$6.19\%) & 0.4007 ($\uparrow$9.69\%)& 2498 ($\uparrow$6.84\%)\\
			TF\_IDF & 0.3344 ($\uparrow$5.05\%) & 0.2352 ($\uparrow$4.02\%) & 0.4480 ($\downarrow$1.79\%) & 0.4260 ($\uparrow$6.23\%) & 0.4000 ($\uparrow$7.90\%)& 2489 ($\uparrow$6.37\%)\\
			BM25 & 0.3316 ($\uparrow$4.84\%) & 0.2343 ($\uparrow$4.88\%) &  0.4420 ($\downarrow$4.07\%) & 0.4250 ($\uparrow$7.05\%) & 0.3947 ($\uparrow$7.84\%) &  2499 ($\uparrow$6.66\%)\\
			\hline
	\end{tabular}}
\end{table}

Table \ref{QE using BingDuckDuckGo} shows the comparative results for the combined approach involving Bing and DuckDuckGo, called  Bing-DuckDuckGo-based QE (BDQE). The BDQE improved the MAP up to 17.53\% and GM\_MAP up to 22.54\%. The BDQE technique shows better retrieval effectiveness in comparison with the DQE technique. However, it failed to improve the retrieval effectiveness when compared to other combined QE techniques. 

\begin{table}[!h]
	\centering
	\caption{Comparison of QE using Bing-DuckDuckGo-based QE (BDQE) on popular models with top 15 expansion terms on the FIRE Dataset. The best result for each method have been highlighted in bold. \label{QE using BingDuckDuckGo}}{
		\begin{tabular}{ |p{1.4cm}||p{1.5cm}|p{1.5cm}|p{1.5cm}|p{1.5cm}|p{1.5cm}|p{1.5cm}|  }
			\hline
			\multicolumn{7}{|c|}{\textbf{Model Performance Without Query Expansion}} \\
			\hline
			Method & MAP & GM\_MAP & P@10 & P@20 & P@30 & \#rel\_ret \\
			\hline
			IFB2  & 0.2765 & 0.1907 & 0.3660 & 0.3560 & 0.3420 & 2330\\
			LGD & 0.2909 & 0.1974 & 0.4100 & 0.3710 & 0.3420 & 2309\\
			I(n)L2& 0.2979 & 0.2023 & 0.4280 & 0.3900 & 0.3553 &  2322\\
			DPH & 0.3133 & 0.2219 & 0.4540 & 0.4040 & 0.3653 & 2338\\
			TF\_IDF & 0.3183 & 0.2261 & 0.4560 & 0.4010 & 0.3707 & 2340\\
			BM25 & 0.3163 & 0.2234 & 0.4600 & 0.3970 & 0.3660 &  2343\\
			\hline \hline
			\multicolumn{7}{| c |}{\textbf{Model Performance With QE using Bing-DuckDuckGo-based QE}}\\ \hline
			IFB2  & 0.3221 ($\uparrow$16.49\%) & 0.2278 ($\uparrow$19.45\%) & 0.4540 \textbf{($\uparrow$24.04\%)} & 0.4200 ($\uparrow$17.98\%) & 0.3913 ($\uparrow$14.42\%) & 2478 ($\uparrow$6.35\%)\\
			
			LGD & 0.3419 \textbf{($\uparrow$17.53\%)} & 0.2419 \textbf{($\uparrow$22.54\%)} & 0.4680 ($\uparrow$14.15\%) & 0.4290 \textbf{($\uparrow$25.81\%)} & 0.4100 \textbf{($\uparrow$19.88\%)} & 2465 \textbf{($\uparrow$6.76\%)}\\
			I(n)L2 & 0.3428 ($\uparrow$15.07\%) & 0.2448 ($\uparrow$21.01\%) &  0.4540 ($\uparrow$6.07\%) & 0.4280 ($\uparrow$9.74\%) & 0.4040 ($\uparrow$13.71\%) &  2475 ($\uparrow$6.59\%)\\
			DPH & 0.3481 ($\uparrow$11.11\%) & 0.2505 ($\uparrow$12.89\%) & 0.4620 ($\uparrow$1.76\%) & 0.4380 ($\uparrow$8.42\%) & 0.4207 ($\uparrow$15.17\%)& 2495 ($\uparrow$6.72\%)\\
			TF\_IDF & 0.3410 ($\uparrow$7.13\%) & 0.2456 ($\uparrow$8.62\%) & 0.4660 ($\uparrow$2.19\%) & 0.4360 ($\uparrow$8.73\%) & 0.4080 ($\uparrow$10.06\%)& 2476 ($\uparrow$5.81\%)\\
			BM25 & 0.3382 ($\uparrow$6.92\%) & 0.2431 ($\uparrow$8.82\%) &  0.4680 ($\uparrow$1.74\%) & 0.4320 ($\uparrow$8.82\%) & 0.4020 ($\uparrow$9.84\%) &  2470 ($\uparrow$5.42\%)\\
			\hline
	\end{tabular}}
\end{table}

Table \ref{QE using GoogleDuckDuckGo} shows the comparative analysis for the combined approach involving Google and DuckDuckGo, called Google-DuckDuckGo-based QE technique (GDQE), on different weighting models with the top 15 expansion terms. The GDQE approach improved the MAP up to 21.12\% and GM\_MAP up to 26.11\%. This proposed approach showed better retrieval effectiveness when compared to DQE and BDQE. 

\begin{table}[!h]
	\centering
	\caption{Comparison of QE using Google-DuckDuckGo-based QE (GDQE) on popular models with top 15 expansion terms on the FIRE Dataset. The best result for each method have been highlighted in bold. \label{QE using GoogleDuckDuckGo}}{
		\begin{tabular}{ |p{1.4cm}||p{1.5cm}|p{1.5cm}|p{1.5cm}|p{1.5cm}|p{1.5cm}|p{1.5cm}|  }
			\hline
			\multicolumn{7}{|c|}{\textbf{Model Performance Without Query Expansion}} \\
			\hline
			Method & MAP & GM\_MAP & P@10 & P@20 & P@30 & \#rel\_ret \\
			\hline
			IFB2  & 0.2765 & 0.1907 & 0.3660 & 0.3560 & 0.3420 & 2330\\
			LGD & 0.2909 & 0.1974 & 0.4100 & 0.3710 & 0.3420 & 2309\\
			I(n)L2& 0.2979 & 0.2023 & 0.4280 & 0.3900 & 0.3553 &  2322\\
			DPH & 0.3133 & 0.2219 & 0.4540 & 0.4040 & 0.3653 & 2338\\
			TF\_IDF & 0.3183 & 0.2261 & 0.4560 & 0.4010 & 0.3707 & 2340\\
			BM25 & 0.3163 & 0.2234 & 0.4600 & 0.3970 & 0.3660 &  2343\\
			\hline \hline
			\multicolumn{7}{| c |}{\textbf{Model Performance With QE using Google-DuckDuckGo-based QE}}\\ \hline
			IFB2  & 0.3349 \textbf{($\uparrow$21.12\%)} & 0.2405 \textbf{($\uparrow$26.11\%)} & 0.4580 \textbf{($\uparrow$25.14\%)} & 0.4310 \textbf{($\uparrow$21.07\%)} & 0.4067 \textbf{($\uparrow$18.92\%)} & 2478 ($\uparrow$6.48\%)\\
			
			LGD & 0.3482 ($\uparrow$19.70\%) & 0.2472 ($\uparrow$25.23\%) & 0.4580 ($\uparrow$11.71\%) & 0.4350 ($\uparrow$17.25\%) & 0.4053 ($\uparrow$18.51\%) & 2480 ($\uparrow$7.40\%)\\
			I(n)L2 & 0.3474 ($\uparrow$16.61\%) & 0.2460 ($\uparrow$21.60\%) &  0.4600 ($\uparrow$7.48\%) & 0.4260 ($\uparrow$9.23\%) & 0.4053 ($\uparrow$14.07\%) &  2494 \textbf{($\uparrow$7.41\%)}\\
			DPH & 0.3557 ($\uparrow$13.53\%) & 0.2581 ($\uparrow$16.31\%) & 0.4780 ($\uparrow$5.28\%) & 0.4400 ($\uparrow$8.91\%) & 0.4193 ($\uparrow$14.78\%)& 2503 ($\uparrow$7.06\%)\\
			TF\_IDF & 0.3495 ($\uparrow$9.80\%) & 0.2524 ($\uparrow$11.63\%) & 0.4720 ($\uparrow$3.51\%) & 0.4220 ($\uparrow$5.24\%) & 0.4147 ($\uparrow$11.87\%)& 2497 ($\uparrow$6.71\%)\\
			BM25 & 0.3467 ($\uparrow$9.61\%) & 0.2499 ($\uparrow$11.86\%) &  0.4680 ($\uparrow$1.74\%) & 0.4270 ($\uparrow$7.56\%) & 0.4087 ($\uparrow$11.67\%) &  2495 ($\uparrow$6.49\%)\\
			\hline
	\end{tabular}}
\end{table}

Table  \ref{QE using GoogleBing} shows the corresponding results for the combined approach involving Google and Bing,  called Google-Bing-based QE technique (GBQE). It improved the MAP up to 22.64\% and GM\_MAP up to 29.16\%. This GBQE technique showed better retrieval effectiveness in comparison to all the proposed techniques, except GBDQE. 

\begin{table}[!h]
	\centering
	\caption{Comparison of QE using Google-Bing-based QE (GBQE) on popular models with top 15 expansion terms on the FIRE Dataset. The best result for each method have been highlighted in bold. \label{QE using GoogleBing}}{
		\begin{tabular}{ |p{1.4cm}||p{1.5cm}|p{1.5cm}|p{1.5cm}|p{1.5cm}|p{1.5cm}|p{1.5cm}|  }
			\hline
			\multicolumn{7}{|c|}{\textbf{Model Performance Without Query Expansion}} \\
			\hline
			Method & MAP & GM\_MAP & P@10 & P@20 & P@30 & \#rel\_ret \\
			\hline
			IFB2  & 0.2765 & 0.1907 & 0.3660 & 0.3560 & 0.3420 & 2330\\
			LGD & 0.2909 & 0.1974 & 0.4100 & 0.3710 & 0.3420 & 2309\\
			I(n)L2& 0.2979 & 0.2023 & 0.4280 & 0.3900 & 0.3553 &  2322\\
			DPH & 0.3133 & 0.2219 & 0.4540 & 0.4040 & 0.3653 & 2338\\
			TF\_IDF & 0.3183 & 0.2261 & 0.4560 & 0.4010 & 0.3707 & 2340\\
			BM25 & 0.3163 & 0.2234 & 0.4600 & 0.3970 & 0.3660 &  2343\\
			\hline \hline
			\multicolumn{7}{| c |}{\textbf{Model Performance With QE using Google-Bing-based QE}}\\ \hline
			IFB2  & 0.3391 \textbf{($\uparrow$22.64\%)} & 0.2463 \textbf{($\uparrow$29.16\%)} & 0.4600 \textbf{($\uparrow$25.68\%)} & 0.4260 \textbf{($\uparrow$19.66\%)} & 0.4065 ($\uparrow$18.86\%) & 2561 ($\uparrow$9.91\%)\\
			
			LGD & 0.3539 ($\uparrow$21.66\%) & 0.2535 ($\uparrow$28.41\%) & 0.4700 ($\uparrow$14.63\%) & 0.4360 ($\uparrow$17.52\%) & 0.4092 \textbf{($\uparrow$19.65\%)} & 2559 \textbf{($\uparrow$10.83\%)}\\
			I(n)L2 & 0.3501 ($\uparrow$17.52\%) & 0.2522 ($\uparrow$24.67\%) &  0.4790 ($\uparrow$11.92\%) & 0.4291 ($\uparrow$10.02\%) & 0.4021 ($\uparrow$13.17\%) &  2568 ($\uparrow$10.59\%)\\
			DPH & 0.3631 ($\uparrow$15.90\%) & 0.2672 ($\uparrow$20.41\%) & 0.4832 ($\uparrow$6.43\%) & 0.4429 ($\uparrow$9.63\%) & 0.4227 ($\uparrow$15.71\%)& 2575 ($\uparrow$9.90\%)\\
			TF\_IDF & 0.3577 ($\uparrow$12.38\%) & 0.2635 ($\uparrow$16.54\%) & 0.4595 ($\uparrow$0.77\%) & 0.4372 ($\uparrow$9.03\%) & 0.4142 ($\uparrow$11.73\%)& 2562 ($\uparrow$9.49\%)\\
			BM25 & 0.3554 ($\uparrow$12.36\%) & 0.2621 ($\uparrow$17.32\%) &  0.4640 ($\uparrow$0.87\%) & 0.4359 ($\uparrow$9.80\%) & 0.4071 ($\uparrow$11.23\%) &  2559 ($\uparrow$9.22\%)\\
			\hline
	\end{tabular}}
\end{table}

Table \ref{QE using GoogleBingDuckDuckGo} shows the comparative analysis of the query expansion approach by combining all the three search engines, called Google-Bing-DuckDuckGo-based QE technique (GBDQE). It improved the MAP up to 25.89\% and GM\_MAP up to 30.83\%. The proposed GBDQE approach showed the best retrieval performance in comparison to all the other proposed expansion techniques.

\begin{table}[!h]
	\centering
	\caption{Comparison of QE using Google-Bing-DuckDuckGo-based QE (GBDQE) on popular models with top 15 expansion terms on the FIRE Dataset. The best result for each method have been highlighted in bold. \label{QE using GoogleBingDuckDuckGo}}{
		\begin{tabular}{ |p{1.4cm}||p{1.5cm}|p{1.5cm}|p{1.5cm}|p{1.5cm}|p{1.5cm}|p{1.5cm}|  }
			\hline
			\multicolumn{7}{|c|}{\textbf{Model Performance Without Query Expansion}} \\
			\hline
			Method & MAP & GM\_MAP & P@10 & P@20 & P@30 & \#rel\_ret \\
			\hline
			IFB2  & 0.2765 & 0.1907 & 0.3660 & 0.3560 & 0.3420 & 2330\\
			LGD & 0.2909 & 0.1974 & 0.4100 & 0.3710 & 0.3420 & 2309\\
			I(n)L2& 0.2979 & 0.2023 & 0.4280 & 0.3900 & 0.3553 &  2322\\
			DPH & 0.3133 & 0.2219 & 0.4540 & 0.4040 & 0.3653 & 2338\\
			TF\_IDF & 0.3183 & 0.2261 & 0.4560 & 0.4010 & 0.3707 & 2340\\
			BM25 & 0.3163 & 0.2234 & 0.4600 & 0.3970 & 0.3660 &  2343\\
			\hline \hline
			\multicolumn{7}{| c |}{\textbf{Model Performance With QE using Google-Bing-DuckDuckGo-based QE}}\\ \hline
			IFB2  & 0.3481 \textbf{($\uparrow$25.89\%)} & 0.2495 \textbf{($\uparrow$30.83\%)} & 0.4610 \textbf{($\uparrow$25.96\%)} & 0.4289 \textbf{($\uparrow$20.48\%)} & 0.4092 ($\uparrow$19.65\%) & 2591 \textbf{($\uparrow$11.20\%)}\\
			
			LGD & 0.3552 ($\uparrow$22.10\%) & 0.2539 ($\uparrow$28.62\%) & 0.4713 ($\uparrow$14.95\%) & 0.4392 ($\uparrow$18.38\%) & 0.4099 \textbf{($\uparrow$19.85\%)} & 2563 ($\uparrow$11.11\%)\\
			I(n)L2 & 0.3519 ($\uparrow$18.13\%) & 0.2531 ($\uparrow$25.11\%) &  0.4810 ($\uparrow$12.38\%) & 0.4309 ($\uparrow$10.49\%) & 0.4052 ($\uparrow$14.04\%) &  2572 ($\uparrow$10.76\%)\\
			DPH & 0.3640 ($\uparrow$16.18\%) & 0.2681 ($\uparrow$20.82\%) & 0.4841 ($\uparrow$6.63\%) & 0.4438 ($\uparrow$9.85\%) & 0.4238 ($\uparrow$16.01\%)& 2581 ($\uparrow$10.39\%)\\
			TF\_IDF & 0.3583 ($\uparrow$12.57\%) & 0.2649 ($\uparrow$17.16\%) & 0.4611 ($\uparrow$1.11\%) & 0.4381 ($\uparrow$9.25\%) & 0.4157 ($\uparrow$12.14\%)& 2566 ($\uparrow$9.66\%)\\
			BM25 & 0.3569 ($\uparrow$12.84\%) & 0.2638 ($\uparrow$18.08\%) &  0.4651 ($\uparrow$1.10\%) & 0.4367 ($\uparrow$10.00\%) & 0.4082 ($\uparrow$11.53\%) &  2568 ($\uparrow$9.60\%)\\
			\hline
	\end{tabular}}
\end{table}

\begin{figure}[!h]
	\centering 
	\includegraphics[width=12cm, height=7.5cm]{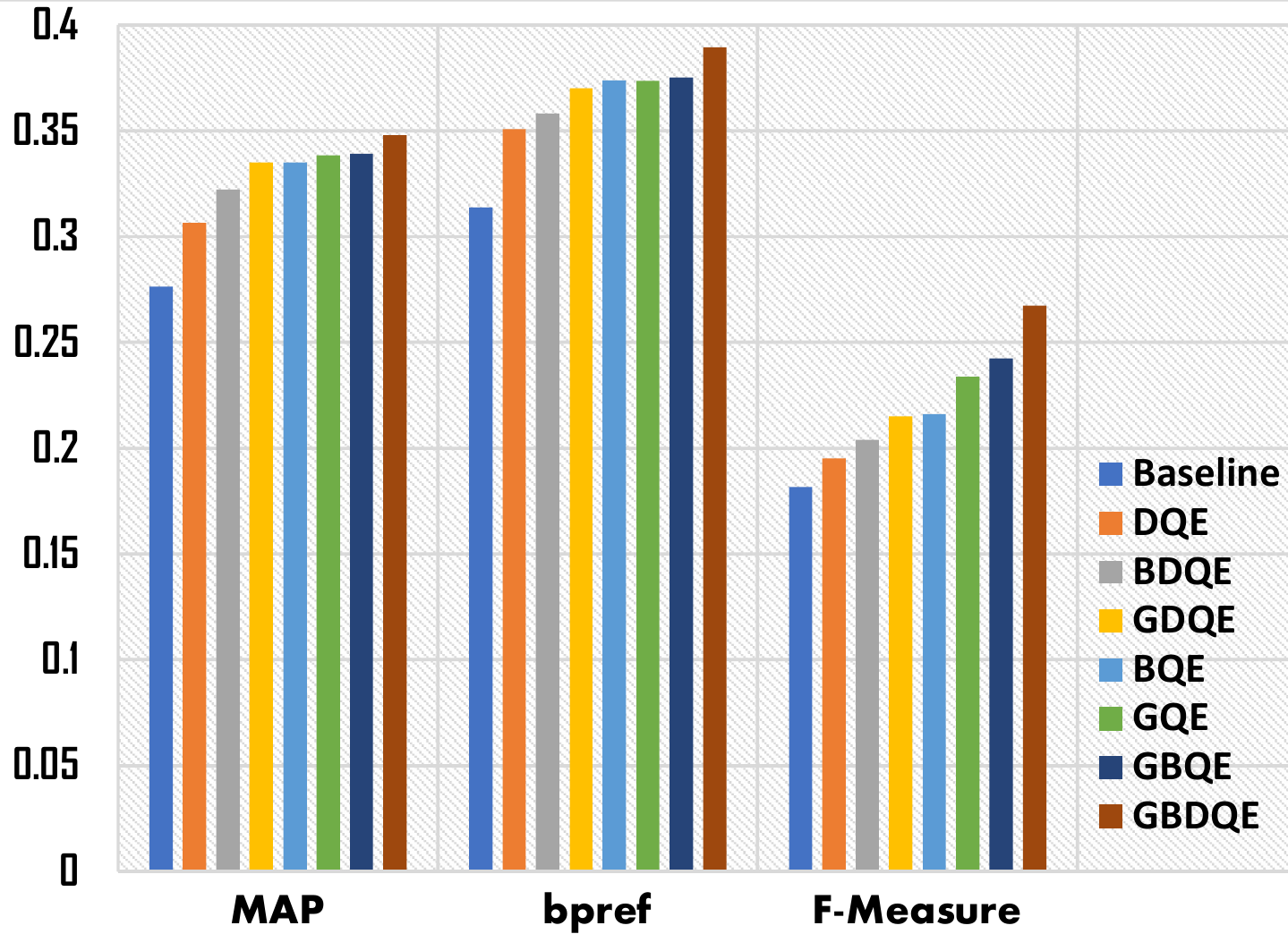}  
	\caption{Comparative analysis of WKQE techniques with baseline.}
	\label{Comparision btwn WKQE} 
\end{figure}

Figure \ref{Comparision btwn WKQE} compares all the proposed WKQE techniques in terms of MAP, bpref, and F-Measure with baseline (unexpanded query). IFB2 weighting model was used here as the baseline. It can be observed that all WKQE techniques achieved a significant improvement over baseline, while in particular GBDQE  outperformed other WKQE techniques. GBDQE improved the MAP by 25.89\%, bpref by 24.09\%, and F-measure by 47.25\% over the baseline on the FIRE dataset. Being the best case, we have chosen the GBDQE technique for comparative analysis with other state-of-the-art approaches. 

\begin{figure}[!h]
	\centering 
	\includegraphics[width=12cm, height=7.5cm]{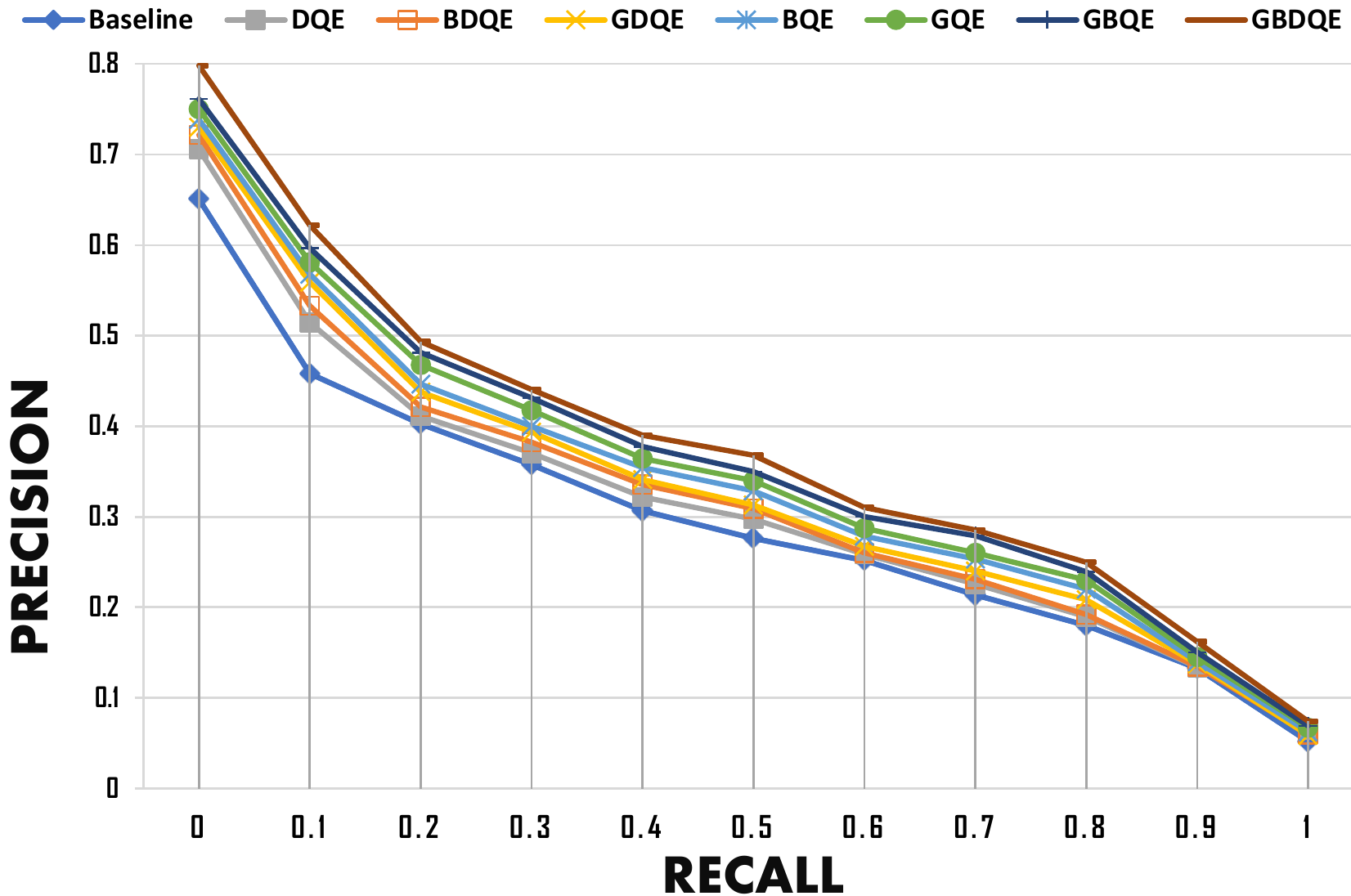}  
	\caption{Comparative analysis of precision-recall curve of the proposed WKQE techniques with baseline.}
	\label{P-R curve_WKQE} 
\end{figure}

Figure \ref{P-R curve_WKQE} shows the comparative analysis of the precision-recall curve of WKQE techniques with the baseline (i.e., unexpanded query). IFB2 weighting model was used here as baseline. This graph plots the interpolated precision of an IR system using 11 standard cut-off values from the recall levels, i.e, \{0, 0.1, 0.2, 0.3, ...,1.0\}. These graphs showing average plot of retrieval results are widely used to evaluate IR systems that return ranked documents. Comparisons are best made in three different recall ranges: 0 to 0.2, 0.2 to 0.8, and 0.8 to 1. These ranges characterize high precision, middle recall, and high recall performance respectively. We can see that the proposed WKQE techniques show significant improvement over the baseline.

\begin{figure}[h]%
	\centering
	\subfloat[]{{\includegraphics[width=4.5cm]{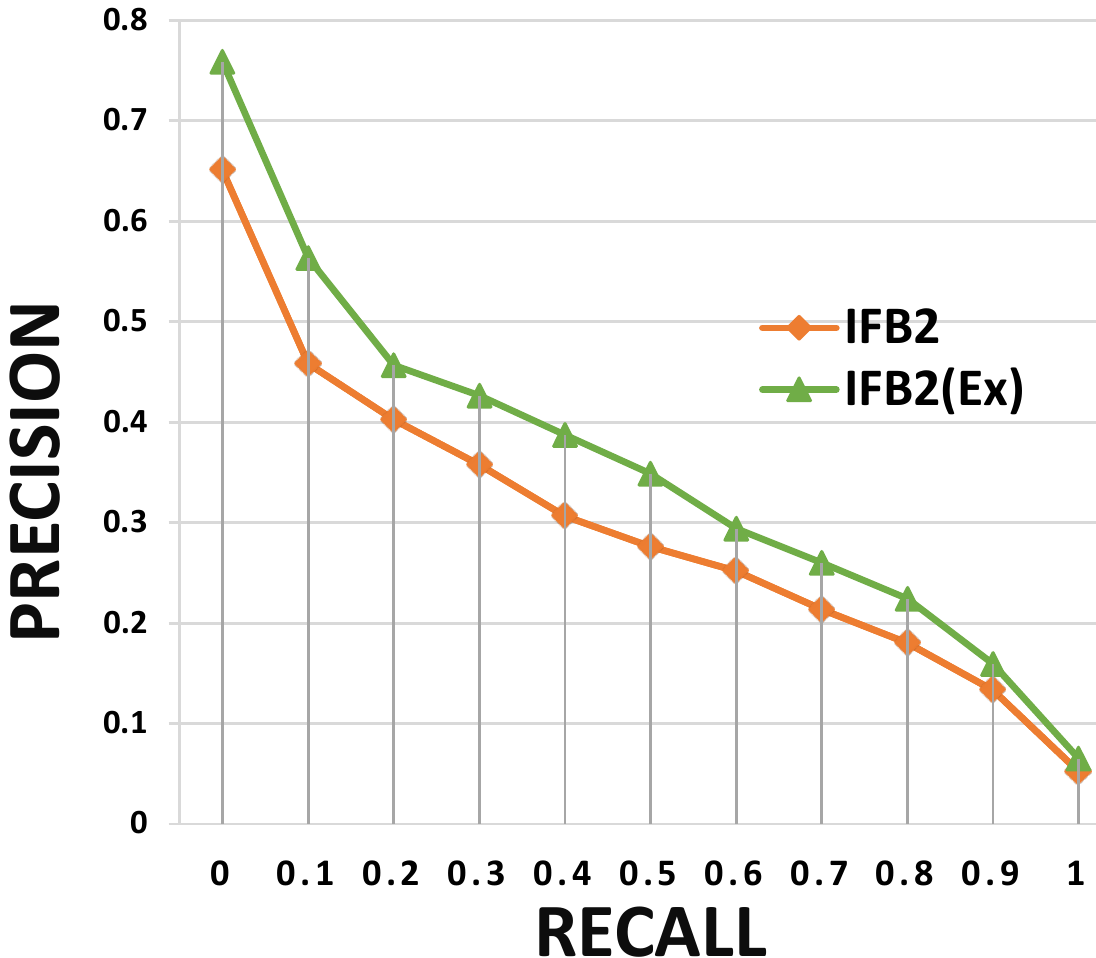} \label{fig:p-r curve IFB2} }}%
	\subfloat[]{{\includegraphics[width=4.5cm]{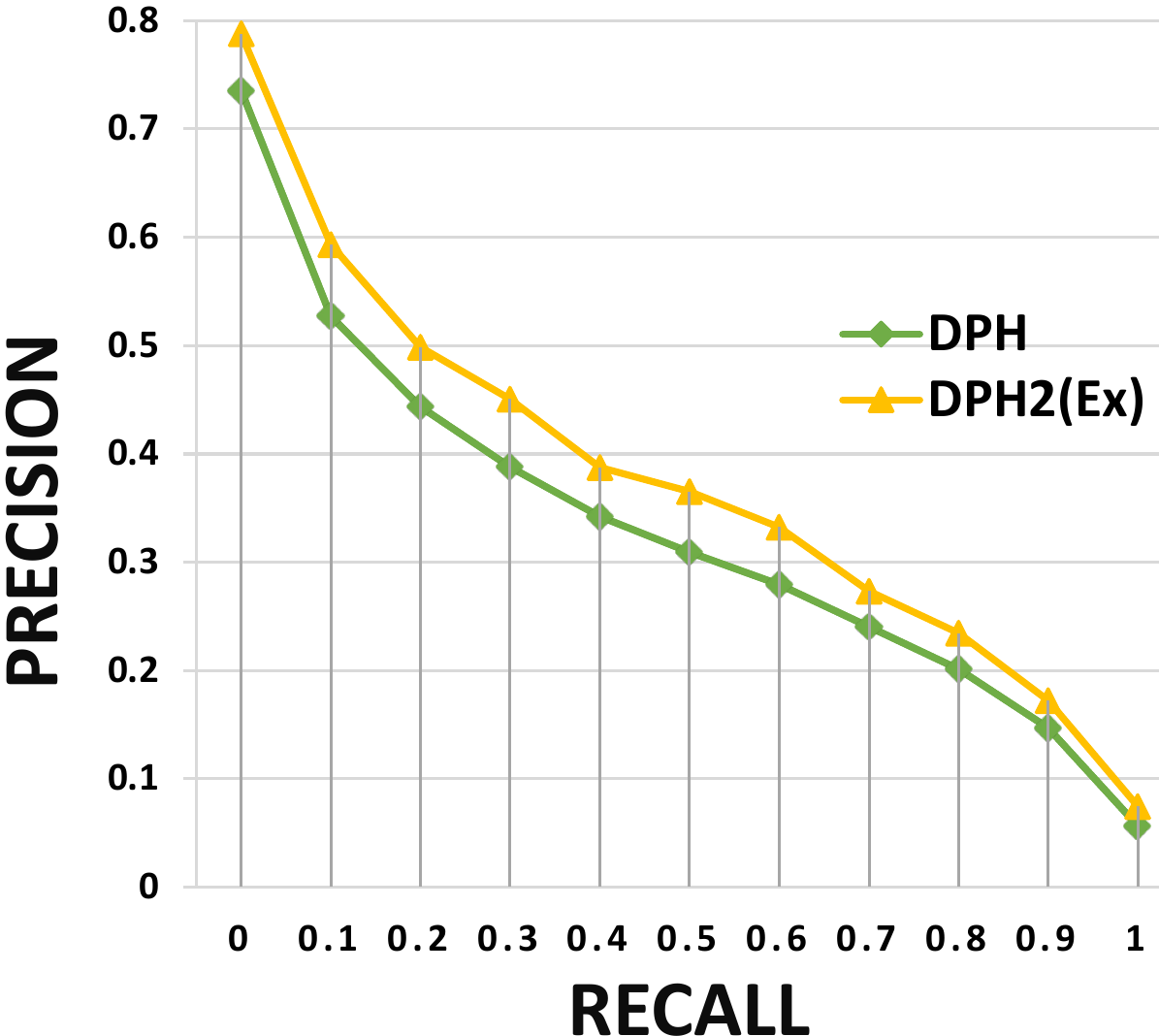} \label{fig:p-r curve InL2} }}%
	\subfloat[]{{\includegraphics[width=4.5cm]{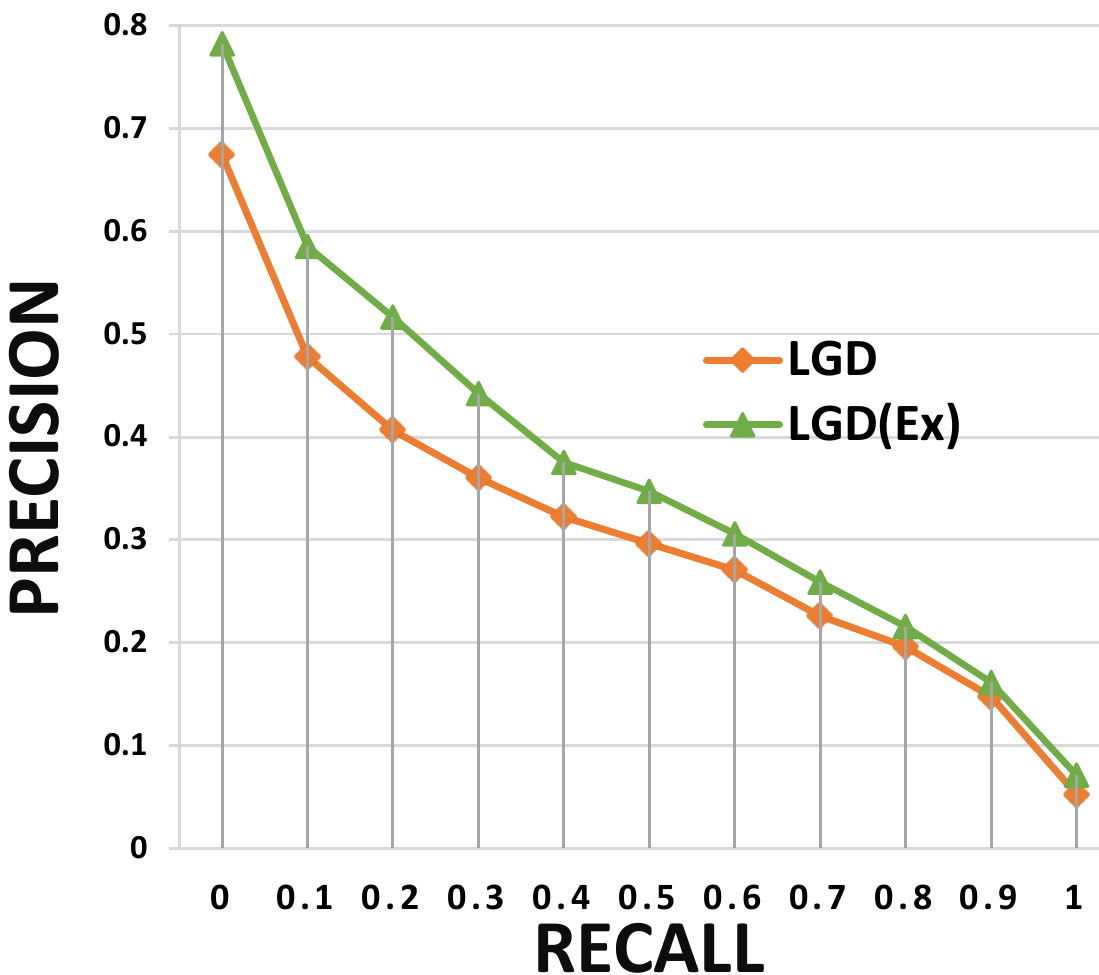}  \label{fig:p-r curve LGD} }}
	\qquad
	\subfloat[]{{\includegraphics[width=4.5cm]{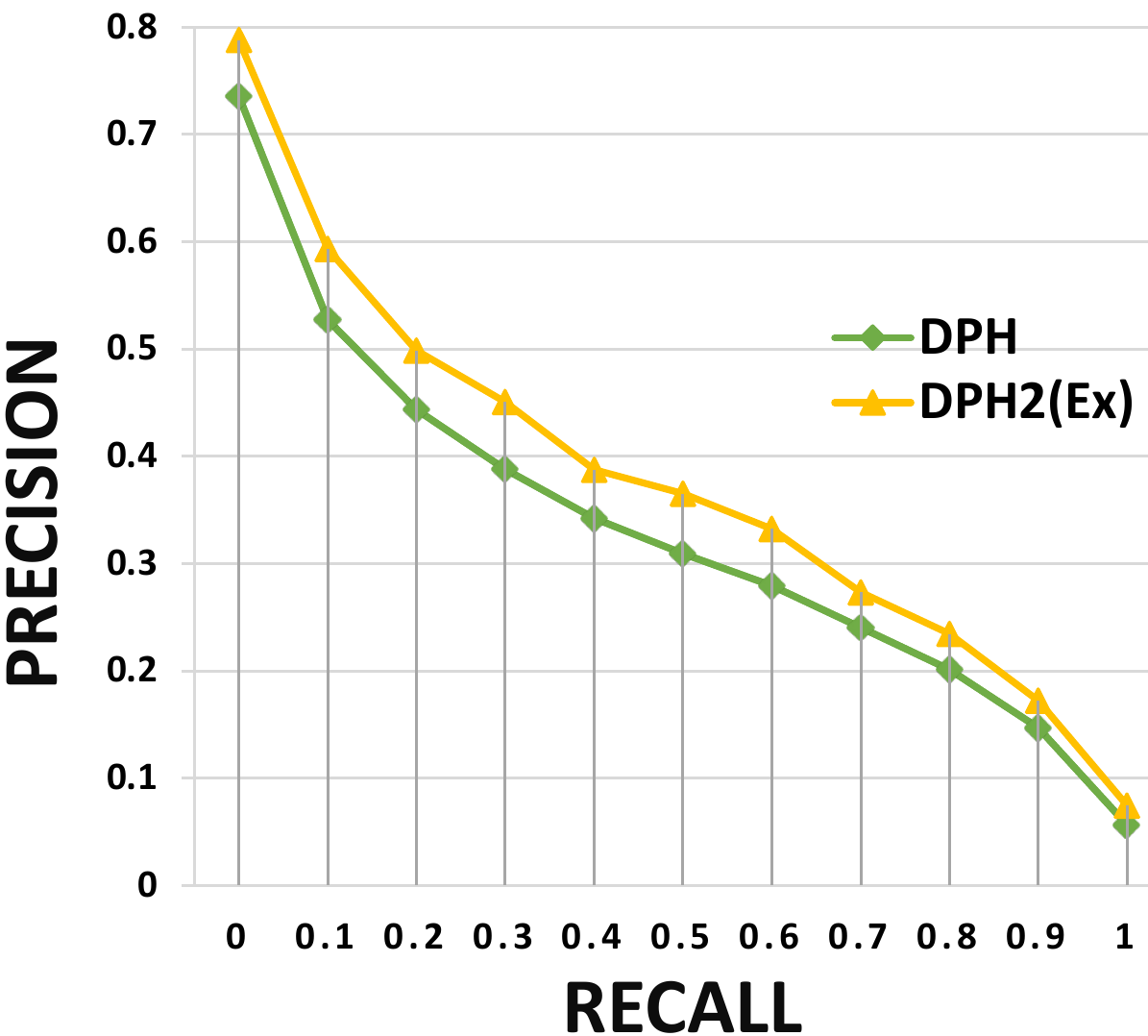}  \label{fig:p-r curve DPH} }}%
	\subfloat[]{{\includegraphics[width=4.5cm]{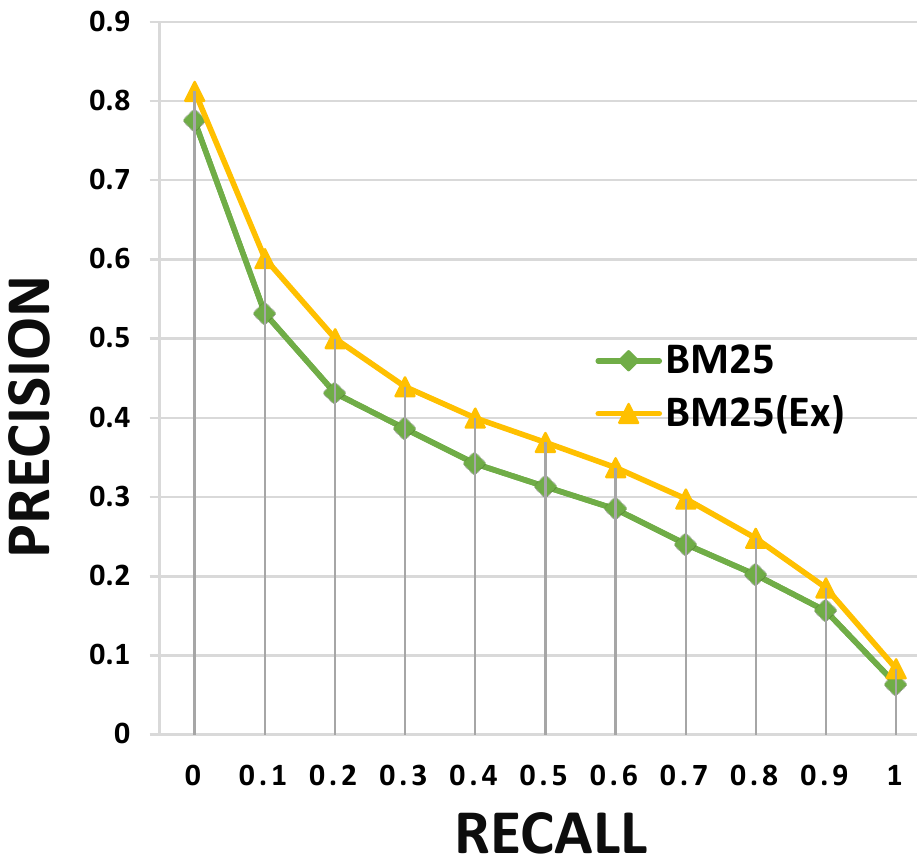} \label{fig:p-r curve BM25} }}%
	\subfloat[]{{\includegraphics[width=4.5cm]{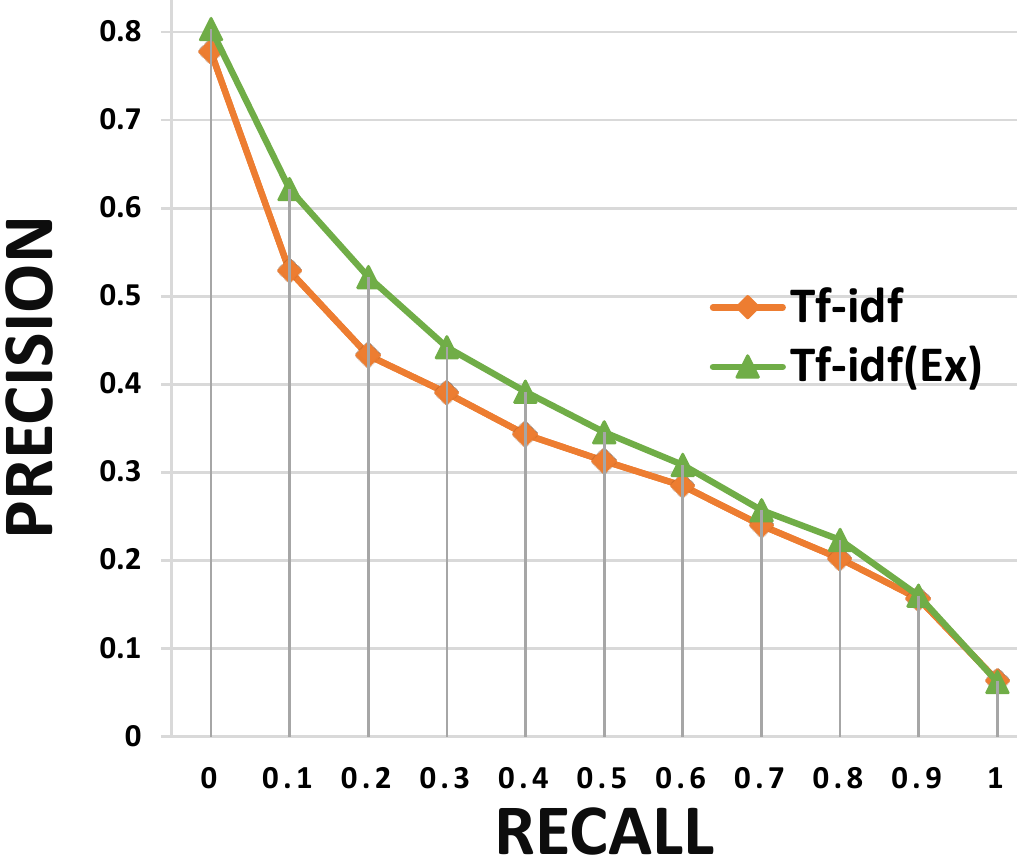} \label{fig:p-r curve TFIDF} }}
	\caption{Comparative analysis of the Precision-Recall curve of GBDQE technique using popular weighting models on the FIRE dataset. In the legend, (Ex) denotes the performance with query expansion, while no parenthesis denotes unexpanded query.}%
	\label{fig:Comparative analysis of Precision-Recall curve individually}%
\end{figure}

Graphs in Fig. \ref{fig:Comparative analysis of Precision-Recall curve individually} show the improvement in retrieval results of GBDQE technique when compared with the initial unexpanded queries.  As indicated in the graph legend, (Ex) denotes the performance with query expansion, while no parenthesis denotes unexpanded query. The P-R curves show the effectiveness of the proposed GBDQE technique with all the  popular weighting models. Among all the weighting models, the IFB2 weighting model provides the best retrieval performance with the proposed GBDQE technique.

Graphs in the Figure \ref{fig:Comparative analysis of MAP, bpref} compare the GBDQE technique with the unexpanded queries in terms of MAP, bpref, and P@5 with various weighting models on the FIRE dataset. Here, MAP shows the overall performance of the GBDQE technique, P@5 measures the precision over the top 5 documents retrieved, and bpref measures a preference relation about how many documents judged as relevant are ranked before the documents judged as irrelevant.  

\begin{figure}[!h]%
	\centering
	\subfloat[]{{\includegraphics[width=7cm]{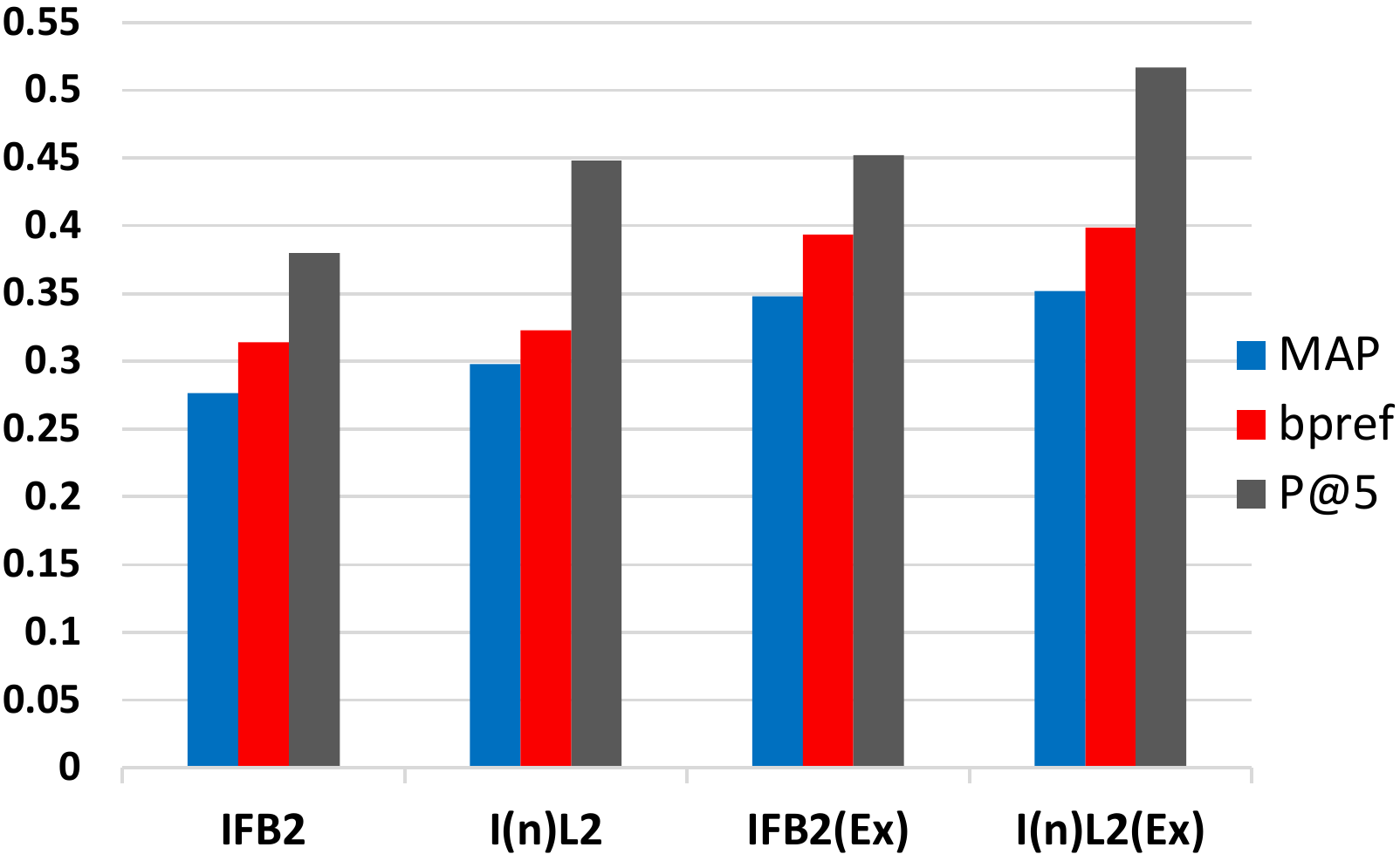}    \label{fig:Chart_IFB2_InL2} }}%
	\subfloat[]{{\includegraphics[width=7cm]{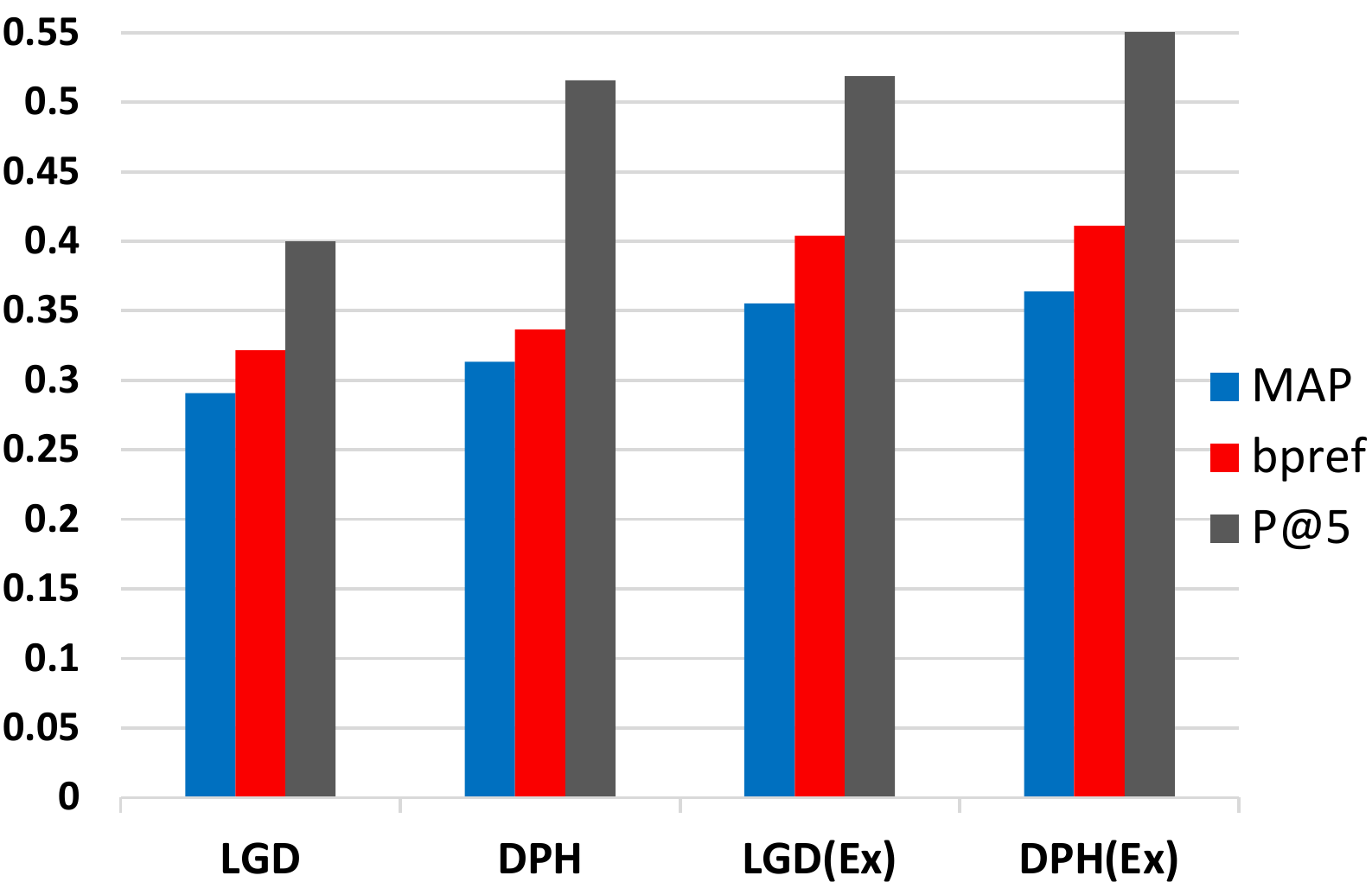}      \label{fig:Chart_LGD_DPH} }}%
	\qquad
	
	\subfloat[]{{\includegraphics[width=7cm]{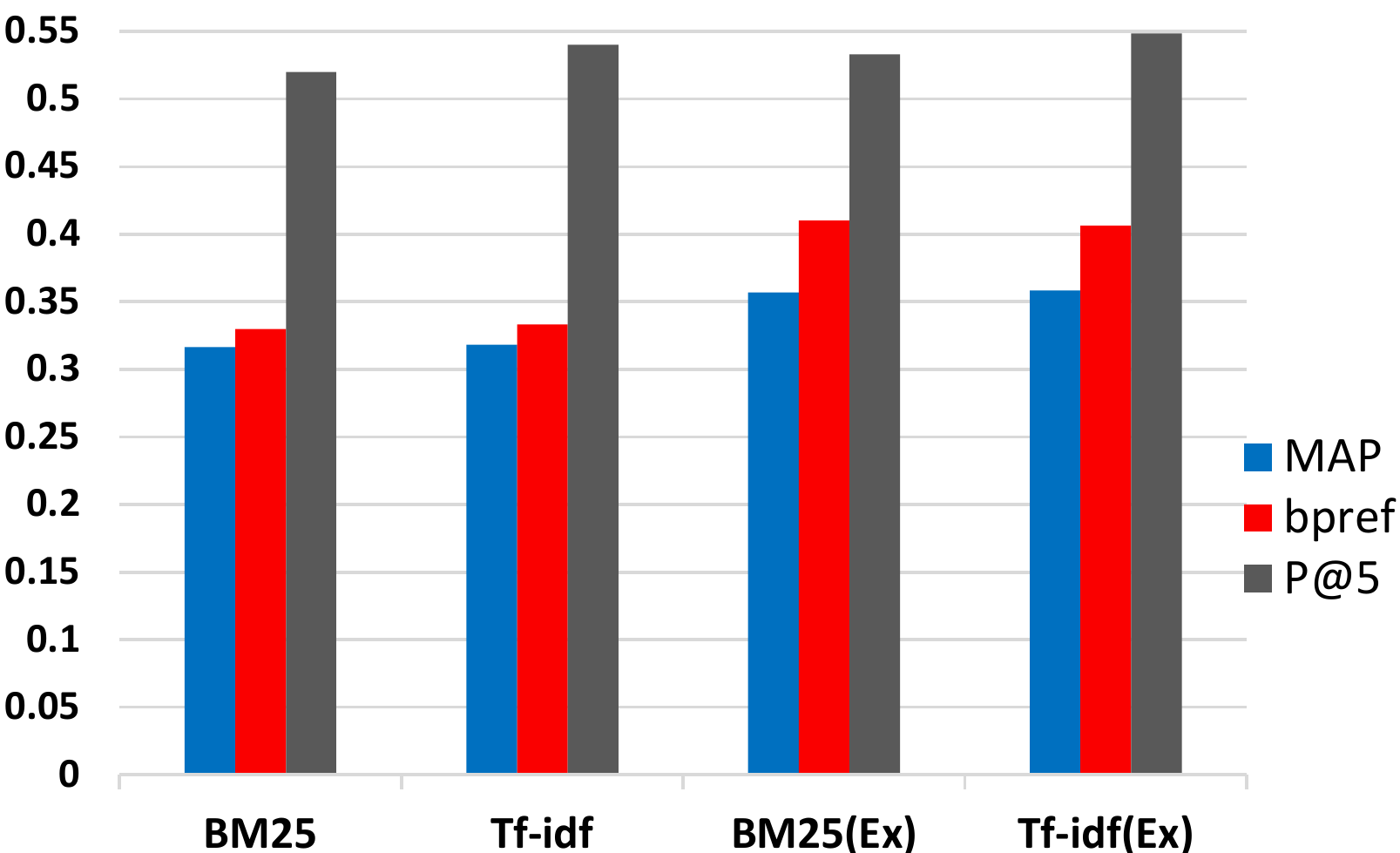}    \label{fig:Chart_BM25_TFIDF} }}
	\caption{Comparative analysis of GBDQE technique in terms of MAP, bpref, and P@5 with various weighting models on the FIRE dataset.}%
	\label{fig:Comparative analysis of MAP, bpref}%
\end{figure} 

After evaluating the performance of the proposed QE technique on several popular evaluation metrics, it can be concluded that the proposed WKQE techniques perform well with popular weighting models on several evaluation parameters. Therefore, the proposed WKQE techniques are effective in improving information retrieval results.

We have also compared our approach with related state-of-the-art works of Parapar et al. \cite{parapar2014score} and Singh and Saran \cite{singh2017term}. 
Parapar et al. \cite{parapar2014score} presented an approach to minimize the number of non-relevant documents in the pseudo-relevant set. These unwanted documents adversely affect the selection of expansion terms. To automatically determine the number of documents to be selected for the pseudo-relevant set for each query, they studied the score distributions in the initial retrieval (i.e., documents retrieved in response to the initial query). The goal of their study was to come-up with a  threshold score to differentiate between relevant and non-relevant documents.  Singh and Saran \cite{singh2017term} method combines the co-occurrence, context window, and semantic similarity based approaches to select the best expansion terms for query expansion. It uses the WordNet-based semantic similarity approach for ranking of expanded terms. The Evaluation of their approach shows a significant improvement over baseline. Finally, their paper suggests the use of context window based query expansion (CWBQE), co-occurrence and semantic based query expansion (CSBQE), and context window and semantic based query expansion (CWSBQE) to improve retrieval effectiveness of an information retrieval system.

Table \ref{Comparision} presents a comparison of the proposed WKQE techniques with Parapar et al. \cite{parapar2014score} and Singh and Saran \cite{singh2017term} models in terms of mean average precision with the top 15 expansion terms on the FIRE dataset. We can observe that the retrieval effectiveness of the proposed GBDQE technique is better than Parapar et al.'s and Singh and Saran's model. Although, Parapar et al.'s and Singh and Saran's models perform well in comparison to the QE using DuckDuckGo alone (DQE). 

\begin{table}[!h]
	\centering
	\caption{Comparative analysis of the proposed WKQE techniques with Parapar et al. and Singh \& Saran's models in terms of MAP values. \label{Comparision}}{
		
		\begin{tabular}{ | M{1.6cm} | M{4.5cm} | M{3.3cm} |  }
			\hline 
			
			\textbf{Data Set} & \textbf{Methods} & \textbf{MAP}  \\ \hline 
			\multirow{6}{2cm}{\centering {FIRE}} & Baseline (IFB2) & 0.2765
			\\\cline{2-3}  & Parapar et al. model \cite{parapar2014score} & 0.3178 (14.74\%)
			\\\cline{2-3}  & Singh and Saran model \cite{singh2017term}  & 0.3286 (18.84\%) 
			\\\cline{2-3}    & GDQE (Proposed) & 0.3349 (21.12\%)
			\\\cline{2-3}    & BQE (Proposed) & 0.3350 (21.16\%)
			\\\cline{2-3}    & GQE (Proposed) & 0.3383 (22.35\%)
			\\\cline{2-3}    & GBQE (Proposed) & 0.3391 (22.64\%)
			\\\cline{2-3}    & \textbf{GBDQE (Proposed)} & \textbf{0.3481 (25.89\%)}
			\\ \hline

	\end{tabular}}
\end{table}

\begin{figure}[!h]
	\centering 
	\includegraphics[width=12cm, height=7cm]{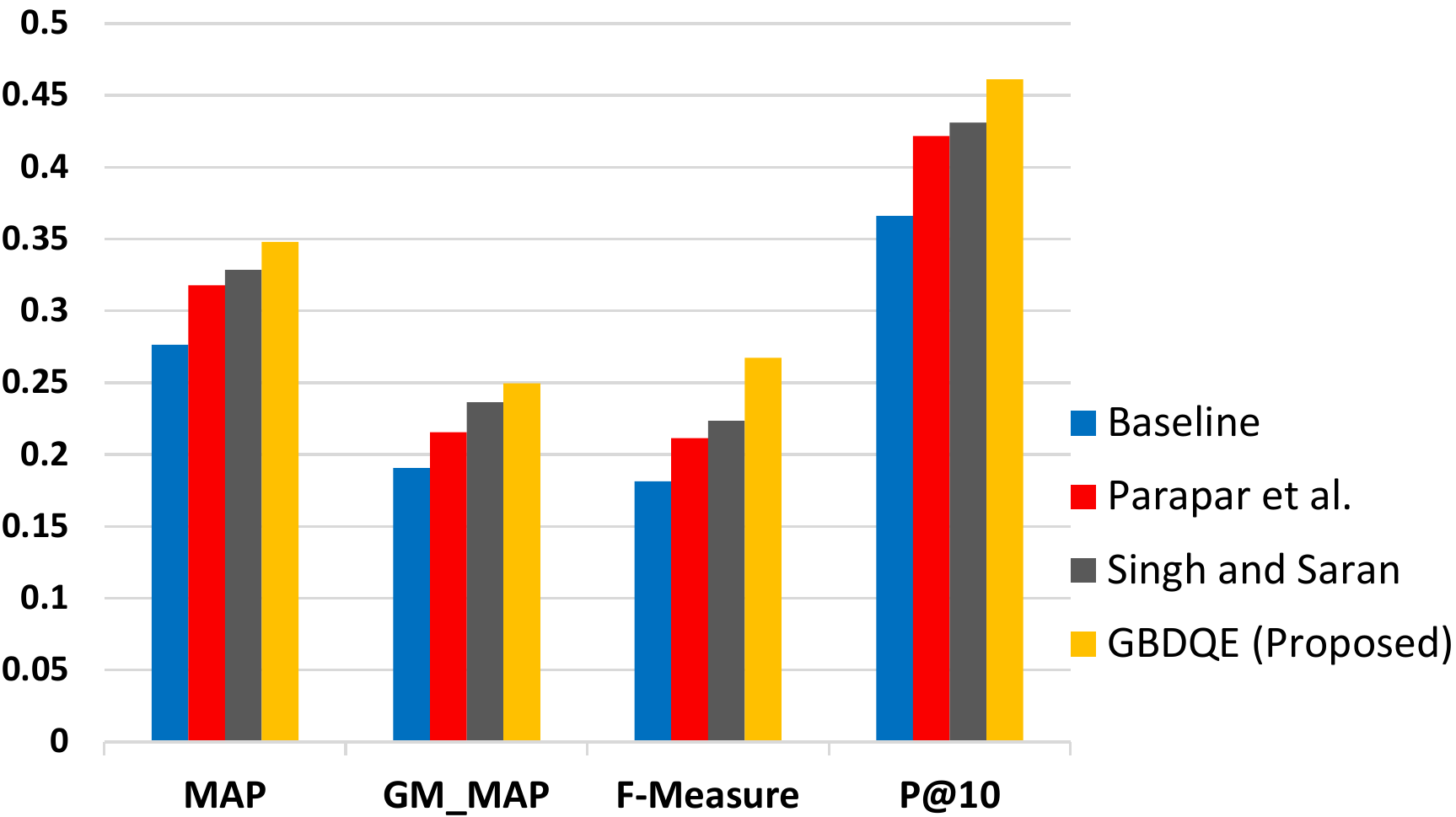}  
	\caption{Comparative analysis of GBDQE technique with baseline and other related approaches.}
	\label{WKQE_all} 
\end{figure}

Figure \ref{WKQE_all} compares the GBDQE technique in terms of MAP, GM\_MAP, F-Measure, and P@10 with baseline (IFB2), Parapar et al.'s, and, Singh and Saran's model. It can be clearly seen that the proposed GBDQE techniques achieves a significant improvement over Parapar et al.'s and Singh and Saran's technique. 

\subsection{Performance variation with the number of pseudo-relevant documents}\label{No of PRD}
Owing to the high density of the relevant documents in the top-ranked documents, it may be intuitive to deduce that a fewer number of top-ranked documents may prove sufficient for query expansion as far as  retrieval performance is concerned. However, as evident from the experimental results shown in Table \ref{Effect of Pseudo-relevant Documents}, this may not always be true.

Table \ref{Effect of Pseudo-relevant Documents} depicts the MAP value of the proposed model for each of the considered weighting methods. It can be clearly observed that for all the considered methods, the retrieval performance of the proposed model increases as the number of pseudo-relevant documents is increased up to 20 (25 for I(n)L2 method). This can be attributed to the fact that the lesser ranked pseudo relevant documents also contain relevant expansion terms. For instance, selecting a very small number of pseudo-relevant documents for Synonymy or Polysemy types of queries may produce very bad results in terms of retrieval performance. This can be attributed to the fact that the documents relevant to these queries may not exist in the considered subset of the pseudo-relevant documents. Moreover, with very few relevant documents, the IR system may not have enough information to extract all possible relevant expansion terms. 

Further increment in the number of pseudo-relevant documents  degrades the performance of the proposed model due to the addition of irrelevant terms from the lower ranked documents. This observation was also made in \cite{roy2019estimating}.  
In summary, choosing 20 pseudo relevant strikes the best balance between choosing relevant and irrelevant expansion terms.

\begin{table}[!h]
	\centering
	\caption{ Effect of the number of Pseudo-relevant Documents on the performance (MAP) of proposed GBDQE technique on the FIRE Dataset. The best result for each method have been highlighted in bold. \label{Effect of Pseudo-relevant Documents}}{
		\begin{tabular}{ |p{1.4cm}||p{1.5cm}|p{1.4cm}|p{1.4cm}|p{1.4cm}|p{1.4cm}|p{1.4cm}|p{1.4cm}|  }
			\hline
			\multicolumn{8}{|c|}{\textbf{Model Performance vs. Pseudo-relevant Documents}} \\
			\hline
			Method & 5 & 10 & 15 & 20 & 25 & 30 & 50 \\
			\hline
			IFB2  & 0.3047 & 0.3168 & 0.3374 & \textbf{0.3481} & 0.3364 & 0.3281 & 0.3102\\
			LGD  & 0.3066 & 0.3170 & 0.3396 & \textbf{0.3552} & 0.3472 & 0.3214 & 0.3113\\
			I(n)L2& 0.3111 & 0.3274 & 0.3401 & 0.3519 & \textbf{0.3521} & 0.3304 & 0.3203  \\
			DPH  & 0.3198 & 0.3311 & 0.3523 & \textbf{0.3640} & 0.3501 & 0.3224 & 0.3174\\
			Tf-idf & 0.3202 & 0.3374 & 0.3501 & \textbf{0.3583} & 0.3498 & 0.3289 & 0.3212\\
			BM25 & 0.3217 & 0.3301 & 0.3472 & \textbf{0.3569} & 0.3507 & 0.3296 &  0.3109\\ \hline 
			
	\end{tabular}}
\end{table}

\subsection{Performance Variation with Number of Expansion Terms}
There are different points of views on the number of expansion terms to be chosen; the number of expansion terms can vary from one-third of the expansion terms to all terms \cite{azad2019query}. Although, it might not be realistic to use all of the expansion terms, a small set of expansion terms is usually better than a large set of expansion terms due to noise reduction \cite{salton1990improving}. A limited number of expansion terms may also be important to reduce the response time, especially for a large corpus. However, several studies observed that the number of expansion terms is of low relevance and it varies from query to query \cite{billerbeck2003query,billerbeck2004questioning,cao2008context}. 

In our experimental results, we show the model performance with top 15 expansion terms because it gives better retrieval performance compared to the other number of terms. We also did experiments by varying the number of expansion terms from 5 to 50 in our proposed model; Table \ref{Effect of expansion terms} shows the corresponding results. The results show that the variation in performance is limited by varying the weighting methods and the number of expansion terms. It can be clearly observed that initially the retrieval performance of the proposed model increases with an increase in the number of expansion terms, the improvement continues up to 15 (20 for LGD) expansion terms, and any subsequent increase in the number of expansion terms adversely affects the retrieval performance. Based on this, selecting top 15 expansion terms seems to give the best results for our proposed approach. 

\begin{table}[!h]
	\centering
	\caption{ Effect of the number of expansion terms on the performance (MAP) of the proposed GBDQE technique on the FIRE Dataset. The best result for each method have been highlighted in bold. \label{Effect of expansion terms}}{
		\begin{tabular}{ |p{1.4cm}||p{1.5cm}|p{1.4cm}|p{1.4cm}|p{1.4cm}|p{1.4cm}|p{1.4cm}|p{1.4cm}|  }
			\hline
			\multicolumn{8}{|c|}{\textbf{Model Performance vs. Expansion terms}} \\
			\hline
			Method & 5 & 10 & 15 & 20 & 25 & 30 & 50 \\
			\hline
			IFB2  & 0.3067 & 0.3330 & \textbf{0.3481} & 0.3361 & 0.3211 & 0.3166 & 0.2983\\
			LGD  & 0.3365 & 0.3423 & 0.3552 & \textbf{0.3561} & 0.3321 & 0.3283 & 0.3141\\
			I(n)L2& 0.3190 & 0.3375 & \textbf{0.3519} & 0.3448 & 0.3302 & 0.3211 & 0.3079 \\
			DPH  & 0.3462 & 0.3576 & \textbf{0.3640} & 0.3559 & 0.3495 & 0.3410 & 0.3321\\
			Tf-idf & 0.3466 & 0.3511 & \textbf{0.3583} & 0.3498 & 0.3376 & 0.3351 & 0.3227\\
			BM25 & 0.3497 & 0.3461 & \textbf{0.3569} & 0.3463 & 0.3351 &  0.3211 & 0.3191\\ \hline 
			
	\end{tabular}}
\end{table}

Table \ref{expansion terms} shows some examples of initial query and the corresponding expansion terms obtained with the top three proposed approaches. It is interesting to note the difference in the set of expansion terms returned by the different approaches. 
\begin{table}[!h]
	\centering
	\caption{Expansion terms obtained with the top three proposed approaches for selected queries on the FIRE dataset \label{expansion terms}}{
		
		\begin{tabular}{ | M{1.3cm} | M{2cm} | M{3.1cm} | M{3.4cm} |M{3.4cm} | }
			\hline

			\textbf{Query ID} & \textbf{Original query} & \textbf{Expansion terms obtained with GQE} & \textbf{Expansion terms obtained with GBQE}  & \textbf{Expansion terms obtained with GBDQE} \\ \hline 
			135 & India's agriculture-friendly central budget &  India, rs, government, farmers, scheme, union, minister, sector, finance,  agricultural, tax, development, etc. & Budget, india, government, rs, tax, agriculture, finance, farmers, union, minister, lakh, rural, sector, etc.  &  Government, tax, finance, agriculture, union, minister, farmers, indian, health, rural, fiscal, etc. \\ \hline
			140 & Search for life and water in space & Space, nasa, earth, scientists, science, ice, surface, moon, image, martian, search, planet, etc. & Life, mars, earth, space, surface, science, planet, search, nasa, liquid, scientists, solar, etc.  & Life, water, mars, earth, space, science, planets, nasa, surface, planet, search, moon, scientists, solar, etc.\\ \hline
			155 & Attack on the Taj in Mumbai & India, attacks,  hotel, november, taj, indian, terrorists, police, people, pakistan, mahal, terror, etc. & Taj, hotel, india, 2008, attacks, november, terrorists, indian, people, mahal, police, terror, palace, etc. & Hotel, india, 2008, november, attack, indian, terrorist, people, mahal, police, palace, terror, etc. \\ \hline
			161 & George Bush 's anti-terrorism operations & War, bush, president, united, people, iraq, states, terrorism, world,  terror, military, american, security, freedom, etc. & War, bush, president, united, military, states, iraq, terrorism, terror, george, american, terrorist, world, security, administration, etc. & War, president, united, states, military, iraq, american, terrorism, world, security, people, terror, policy, etc. \\ \hline
			
	\end{tabular}}
\end{table}	

\section{Conclusion}
\label{Conclusion}
This paper has introduced a novel Web Knowledge based Query Expansion (WKQE) approach that considers expansion terms from the top pseudo-relevant documents collected from different search engines. Although there is no perfect solution for the vocabulary mismatch problem, the proposed WKQE approach is capable to overcome the primary limitations for term-term and term-to-query relationship. To explore the relationship between the query term to the expanded terms, WKQE approach employs a three level weighting strategy to select relevant expansion terms. First, a tf-itf weighting scheme was used to score the individual terms of the web content, then kNN-based cosine similarity was used to identify the k-nearest neighbor expansion terms of the initial query, and lastly  the Correlation score was used to correlate the expansion terms with the whole query. The combination of the web content extracted from three different search engines works well for selecting expansion terms and the proposed WKQE techniques performed well with these terms on several weighting models. It also yielded better results when compared to the baseline and with the other related state-of-the-art methods.  This article also investigated the retrieval performance of the proposed technique with varying number of  pseudo-relevant documents collected from different search engines and expansion terms. The result based on multiple evaluation metrics and popular weighting models on the FIRE dataset demonstrated the effectiveness of our proposed QE technique in the field of information retrieval.

\section*{Acknowledgements}
	Akshay Deepak has been awarded Young Faculty Research Fellowship (YFRF) of Visvesvaraya PhD Programme of Ministry of Electronics \& Information Technology, MeitY, Government of India. In this regard, he would like to acknowledge that this publication is an outcome of the R\&D work undertaken in the project under the Visvesvaraya PhD Scheme of Ministry of Electronics \& Information Technology, Government of India, being implemented by Digital India Corporation (formerly Media Lab Asia).

\bibliography{bibfile}

\begin{thebibliography}{10}
\expandafter\ifx\csname url\endcsname\relax
  \def\url#1{\texttt{#1}}\fi
\expandafter\ifx\csname urlprefix\endcsname\relax\def\urlprefix{URL }\fi
\expandafter\ifx\csname href\endcsname\relax
  \def\href#1#2{#2} \def\path#1{#1}\fi

\bibitem{sta}
Statista, Average number of search terms for online search queries in the
  united states as of august 2017,
  \url{https://www.statista.com/statistics/269740/number-of-search-terms-in-internet-research-in-the-us/}.

\bibitem{key}
Keyword, Query size by country,
  \url{https://www.keyworddiscovery.com/keyword-stats.html}.

\bibitem{azad2019query}
H.~K. Azad, A.~Deepak, Query expansion techniques for information retrieval: a
  survey, Information Processing \& Management 56~(5) (2019) 1698--1735.

\bibitem{lau1999patterns}
T.~Lau, E.~Horvitz, Patterns of search: analyzing and modeling web query
  refinement, in: UM99 User Modeling, Springer, 1999, pp. 119--128.

\bibitem{merigo2018fifty}
J.~M. Merig{\'o}, W.~Pedrycz, R.~Weber, C.~de~la Sotta, Fifty years of
  information sciences: A bibliometric overview, Information Sciences 432
  (2018) 245--268.

\bibitem{furnas1987vocabulary}
G.~W. Furnas, T.~K. Landauer, L.~M. Gomez, S.~T. Dumais, The vocabulary problem
  in human-system communication, Communications of the ACM 30~(11) (1987)
  964--971.

\bibitem{liu2017multi}
J.~Liu, C.~Liu, Y.~Huang, Multi-granularity sequence labeling model for acronym
  expansion identification, Information Sciences 378 (2017) 462--474.

\bibitem{carpineto2012survey}
C.~Carpineto, G.~Romano, A survey of automatic query expansion in information
  retrieval, ACM Computing Surveys (CSUR) 44~(1) (2012) 1.

\bibitem{lucchese2018efficient}
C.~Lucchese, F.~M. Nardini, R.~Perego, R.~Trani, R.~Venturini, Efficient and
  effective query expansion for web search, in: Proceedings of the 27th ACM
  International Conference on Information and Knowledge Management, ACM, 2018,
  pp. 1551--1554.

\bibitem{dalton2014entity}
J.~Dalton, L.~Dietz, J.~Allan, Entity query feature expansion using knowledge
  base links, in: Proceedings of the 37th international ACM SIGIR conference on
  Research \& development in information retrieval, ACM, 2014, pp. 365--374.

\bibitem{bendersky2012effective}
M.~Bendersky, D.~Metzler, W.~B. Croft, Effective query formulation with
  multiple information sources, in: Proceedings of the fifth ACM international
  conference on Web search and data mining, ACM, 2012, pp. 443--452.

\bibitem{yin2009query}
Z.~Yin, M.~Shokouhi, N.~Craswell, Query expansion using external evidence, in:
  European Conference on Information Retrieval, Springer, 2009, pp. 362--374.

\bibitem{almasri2013wikipedia}
M.~ALMasri, C.~Berrut, J.-P. Chevallet, Wikipedia-based semantic query
  enrichment, in: Proceedings of the sixth international workshop on Exploiting
  semantic annotations in information retrieval, ACM, 2013, pp. 5--8.

\bibitem{anand2015empirical}
R.~Anand, A.~Kotov, An empirical comparison of statistical term association
  graphs with dbpedia and conceptnet for query expansion, in: Proceedings of
  the 7th Forum for Information Retrieval Evaluation, ACM, 2015, pp. 27--30.

\bibitem{cui2002probabilistic}
H.~Cui, J.-R. Wen, J.-Y. Nie, W.-Y. Ma, Probabilistic query expansion using
  query logs, in: Proceedings of the 11th international conference on World
  Wide Web, ACM, 2002, pp. 325--332.

\bibitem{dang2010query}
V.~Dang, B.~W. Croft, Query reformulation using anchor text, in: Proceedings of
  the third ACM international conference on Web search and data mining, ACM,
  2010, pp. 41--50.

\bibitem{kraft2004mining}
R.~Kraft, J.~Zien, Mining anchor text for query refinement, in: Proceedings of
  the 13th international conference on World Wide Web, ACM, 2004, pp. 666--674.

\bibitem{lv2011boosting}
Y.~Lv, C.~Zhai, W.~Chen, A boosting approach to improving pseudo-relevance
  feedback, in: Proceedings of the 34th international ACM SIGIR conference on
  Research and development in Information Retrieval, ACM, 2011, pp. 165--174.

\bibitem{xu2009query}
Y.~Xu, G.~J. Jones, B.~Wang, Query dependent pseudo-relevance feedback based on
  wikipedia, in: Proceedings of the 32nd international ACM SIGIR conference on
  Research and development in information retrieval, ACM, 2009, pp. 59--66.

\bibitem{maron1960relevance}
M.~E. Maron, J.~L. Kuhns, On relevance, probabilistic indexing and information
  retrieval, Journal of the ACM (JACM) 7~(3) (1960) 216--244.

\bibitem{rocchio1971relevance}
J.~J. Rocchio, Relevance feedback in information retrieval.

\bibitem{azad2019new}
H.~K. Azad, A.~Deepak, A new approach for query expansion using wikipedia and
  wordnet, Information Sciences 492 (2019) 147--163.

\bibitem{sahami2006web}
M.~Sahami, T.~D. Heilman, A web-based kernel function for measuring the
  similarity of short text snippets, in: Proceedings of the 15th international
  conference on World Wide Web, AcM, 2006, pp. 377--386.

\bibitem{bollegala2007measuring}
D.~Bollegala, Y.~Matsuo, M.~Ishizuka, Measuring semantic similarity between
  words using web search engines., www 7 (2007) 757--766.

\bibitem{riezler2008translating}
S.~Riezler, Y.~Liu, A.~Vasserman, Translating queries into snippets for
  improved query expansion, in: Proceedings of the 22nd International
  Conference on Computational Linguistics-Volume 1, Association for
  Computational Linguistics, 2008, pp. 737--744.

\bibitem{cui2003query}
H.~Cui, J.-R. Wen, J.-Y. Nie, W.-Y. Ma, Query expansion by mining user logs,
  IEEE Transactions on knowledge and data engineering 15~(4) (2003) 829--839.

\bibitem{white2005study}
R.~W. White, I.~Ruthven, J.~M. Jose, A study of factors affecting the utility
  of implicit relevance feedback, in: Proceedings of the 28th annual
  international ACM SIGIR conference on Research and development in information
  retrieval, ACM, 2005, pp. 35--42.

\bibitem{huang2003relevant}
C.-K. Huang, L.-F. Chien, Y.-J. Oyang, Relevant term suggestion in interactive
  web search based on contextual information in query session logs, Journal of
  the Association for Information Science and Technology 54~(7) (2003)
  638--649.

\bibitem{huang2009analyzing}
J.~Huang, E.~N. Efthimiadis, Analyzing and evaluating query reformulation
  strategies in web search logs, in: Proceedings of the 18th ACM conference on
  Information and knowledge management, ACM, 2009, pp. 77--86.

\bibitem{baeza2011extracting}
R.~Baeza-Yates, A.~Tiberi, Extracting semantic relations from query logs, uS
  Patent 7,895,235 (Feb.~22 2011).

\bibitem{xue2004optimizing}
G.-R. Xue, H.-J. Zeng, Z.~Chen, Y.~Yu, W.-Y. Ma, W.~Xi, W.~Fan, Optimizing web
  search using web click-through data, in: Proceedings of the thirteenth ACM
  international conference on Information and knowledge management, ACM, 2004,
  pp. 118--126.

\bibitem{hua2013clickage}
X.-S. Hua, L.~Yang, J.~Wang, J.~Wang, M.~Ye, K.~Wang, Y.~Rui, J.~Li, Clickage:
  towards bridging semantic and intent gaps via mining click logs of search
  engines, in: Proceedings of the 21st ACM international conference on
  Multimedia, ACM, 2013, pp. 243--252.

\bibitem{riezler2007statistical}
S.~Riezler, A.~Vasserman, I.~Tsochantaridis, V.~Mittal, Y.~Liu, Statistical
  machine translation for query expansion in answer retrieval, in: Annual
  Meeting-Association For Computational Linguistics, Vol.~45, 2007, p. 464.

\bibitem{cao2008context}
H.~Cao, D.~Jiang, J.~Pei, Q.~He, Z.~Liao, E.~Chen, H.~Li, Context-aware query
  suggestion by mining click-through and session data, in: Proceedings of the
  14th ACM SIGKDD international conference on Knowledge discovery and data
  mining, ACM, 2008, pp. 875--883.

\bibitem{fitzpatrick1997automatic}
L.~Fitzpatrick, M.~Dent, Automatic feedback using past queries: social
  searching?, in: ACM SIGIR Forum, Vol.~31, ACM, 1997, pp. 306--313.

\bibitem{wang2007learn}
X.~Wang, C.~Zhai, Learn from web search logs to organize search results, in:
  Proceedings of the 30th annual international ACM SIGIR conference on Research
  and development in information retrieval, ACM, 2007, pp. 87--94.

\bibitem{billerbeck2003query}
B.~Billerbeck, F.~Scholer, H.~E. Williams, J.~Zobel, Query expansion using
  associated queries, in: Proceedings of the twelfth international conference
  on Information and knowledge management, ACM, 2003, pp. 2--9.

\bibitem{wang2008mining}
X.~Wang, C.~Zhai, Mining term association patterns from search logs for
  effective query reformulation, in: Proceedings of the 17th ACM conference on
  Information and knowledge management, ACM, 2008, pp. 479--488.

\bibitem{mcbryan1994genvl}
O.~A. McBryan, Genvl and wwww: Tools for taming the web, in: Proceedings of the
  first international world wide web conference, Vol. 341, Geneva, 1994.

\bibitem{arguello2008document}
J.~Arguello, J.~L. Elsas, J.~Callan, J.~G. Carbonell, Document representation
  and query expansion models for blog recommendation., ICWSM 2008~(0) (2008) 1.

\bibitem{li2007improving}
Y.~Li, W.~P.~R. Luk, K.~S.~E. Ho, F.~L.~K. Chung, Improving weak ad-hoc queries
  using wikipedia asexternal corpus, in: Proceedings of the 30th annual
  international ACM SIGIR conference on Research and development in information
  retrieval, ACM, 2007, pp. 797--798.

\bibitem{aggarwal2012query}
N.~Aggarwal, P.~Buitelaar, Query expansion using wikipedia and dbpedia., in:
  CLEF (Online Working Notes/Labs/Workshop), 2012.

\bibitem{al2014wikipedia}
B.~Al-Shboul, S.-H. Myaeng, Wikipedia-based query phrase expansion in patent
  class search, Information retrieval 17~(5-6) (2014) 430--451.

\bibitem{roy2016using}
D.~Roy, D.~Paul, M.~Mitra, U.~Garain, Using word embeddings for automatic query
  expansion, arXiv preprint arXiv:1606.07608.

\bibitem{bai2007using}
J.~Bai, J.-Y. Nie, G.~Cao, H.~Bouchard, Using query contexts in information
  retrieval, in: Proceedings of the 30th annual international ACM SIGIR
  conference on Research and development in information retrieval, ACM, 2007,
  pp. 15--22.

\bibitem{xu1996query}
J.~Xu, W.~B. Croft, Query expansion using local and global document analysis,
  in: Proceedings of the 19th annual international ACM SIGIR conference on
  Research and development in information retrieval, ACM, 1996, pp. 4--11.

\bibitem{sun2006mining}
R.~Sun, C.-H. Ong, T.-S. Chua, Mining dependency relations for query expansion
  in passage retrieval, in: Proceedings of the 29th annual international ACM
  SIGIR conference on Research and development in information retrieval, ACM,
  2006, pp. 382--389.

\bibitem{robertson1996okapi}
S.~E. Robertson, S.~Walker, M.~Beaulieu, M.~Gatford, A.~Payne, Okapi at trec-4,
  Nist Special Publication Sp (1996) 73--96.

\bibitem{amati2002probabilistic}
G.~Amati, C.~J. Van~Rijsbergen, Probabilistic models of information retrieval
  based on measuring the divergence from randomness, ACM Transactions on
  Information Systems (TOIS) 20~(4) (2002) 357--389.

\bibitem{good1965estimation}
I.~Good, The estimation of probabilities: An essay on modern bayesian methods,
  pp. xi-xii (1965).

\bibitem{amati2008fub}
G.~Amati, G.~Amodeo, M.~Bianchi, C.~Gaibisso, G.~Gambosi, Fub, iasi-cnr and
  university of tor vergata at trec 2008 blog track, Tech. rep., FONDAZIONE UGO
  BORDONI ROME (ITALY) (2008).

\bibitem{clinchant2010information}
S.~Clinchant, E.~Gaussier, Information-based models for ad hoc ir, in:
  Proceedings of the 33rd international ACM SIGIR conference on Research and
  development in information retrieval, ACM, 2010, pp. 234--241.

\bibitem{parapar2014score}
J.~Parapar, M.~A. Presedo-Quindimil, {\'A}.~Barreiro, Score distributions for
  pseudo relevance feedback, Information Sciences 273 (2014) 171--181.

\bibitem{singh2017term}
J.~Singh, A.~Sharan, M.~Saini, Term co-occurrence and context window-based
  combined approach for query expansion with the semantic notion of terms,
  International Journal of Web Science 3~(1) (2017) 32--57.

\bibitem{roy2019estimating}
D.~Roy, D.~Ganguly, M.~Mitra, G.~J. Jones, Estimating gaussian mixture models
  in the local neighbourhood of embedded word vectors for query performance
  prediction, Information Processing \& Management 56~(3) (2019) 1026--1045.

\bibitem{salton1990improving}
G.~Salton, C.~Buckley, Improving retrieval performance by relevance feedback,
  Journal of the American society for information science 41~(4) (1990)
  288--297.

\bibitem{billerbeck2004questioning}
B.~Billerbeck, J.~Zobel, Questioning query expansion: An examination of
  behaviour and parameters, in: Proceedings of the 15th Australasian database
  conference-Volume 27, Australian Computer Society, Inc., 2004, pp. 69--76.

\end{thebibliography}

%
%


\end{document}